\documentclass[preprint,prd,tightenlines,nofootinbib,eqsecnum,superscriptaddress]{revtex4-1}

\usepackage{amsmath}
\usepackage{amsfonts}
\usepackage{amssymb}
\usepackage{mathrsfs}
\usepackage{bm}
\usepackage{graphicx}
\usepackage{hyperref}
\usepackage{color}
\usepackage{array}
\usepackage{booktabs}
\usepackage{verbatim}
\usepackage{pagecolor}
\usepackage{dsfont}
\usepackage{booktabs}

\newcolumntype{C}{>{$}c<{$}}
\AtBeginDocument{
\heavyrulewidth=.08em
\lightrulewidth=.05em
\cmidrulewidth=.03em
\belowrulesep=.65ex
\belowbottomsep=0pt
\aboverulesep=.4ex
\abovetopsep=0pt
\cmidrulesep=\doublerulesep
\cmidrulekern=.5em
\defaultaddspace=.5em
}

\definecolor{CiteColor}{rgb}{0,0,0.35}
\definecolor{URLColor}{rgb}{0,0,0.35}
\hypersetup{colorlinks,citecolor=CiteColor,urlcolor=URLColor}

\newcommand{\alext}[1]{\textcolor{black}{#1}}

\newcommand{\beq}{\begin{equation}}
\newcommand{\eeq}{\end{equation}}
\newcommand{\ud}{\mathrm{d}}
\newcommand{\ui}{\mathrm{i}}

\newcommand{\scB}{\mathscr{B}}
\newcommand{\scD}{\mathscr{D}}

\newcommand{\scL}{\mathscr{L}}
\newcommand{\scN}{\mathscr{N}}

\newcommand{\scT}{\mathscr{T}}

\newcommand{\scV}{\mathscr{V}}

\newcommand{\scY}{\mathscr{Y}}

\newcommand{\FF}{\mathbb{F}}

\newcommand{\RR}{\mathbb{R}}

\newcommand{\Lik}{\mathcal{L}_k}

\newcommand{\Lixi}{\mathcal{L}_{\xi}}

\newtheorem{theorem}{Theorem}

\begin{document}

\title{Multipolar Particles in Helically Symmetric Spacetimes}

\author{Paul Ramond}\email{paul.ramond@obspm.fr}
\affiliation{Laboratoire Univers et Th{\'e}ories, Observatoire de Paris, CNRS, Universit{\'e} PSL, Universit{\'e} de Paris, 92190 Meudon, France}

\author{Alexandre Le Tiec}\email{alexandre.letiec@obspm.fr}
\affiliation{Laboratoire Univers et Th{\'e}ories, Observatoire de Paris, CNRS, Universit{\'e} PSL, Universit{\'e} de Paris, 92190 Meudon, France}
\affiliation{Centro Brasileiro de Pesquisas F{\'i}sicas (CBPF), Rio de Janeiro, CEP 22290-180, Brazil}


\date{\today}

\begin{abstract}
We consider a binary system of spinning compact objects with internal structure, moving along an exactly circular orbit, and modelled within the multipolar gravitational skeleton formalism, up to quadrupolar order. We prove that the worldline of each multipolar particle is an integral curve of the helical Killing vector field, and that the 4-velocity, 4-momentum, spin tensor and quadrupole tensor of each particle are Lie-dragged along those worldlines. The geometrical framework developed in this paper paves the way to an extension of the first law of compact-object binary mechanics up to quadrupolar order.
\end{abstract}

\maketitle

\section{Introduction}

\subsection{Motivation}

Ever since September 14, 2015, the direct detections of gravitational waves generated by binary systems of black holes and neutron stars \cite{Ab.al2.16,Ab.al3.17,Ab.al.19} have marked the dawn of a new era in astronomy. Over the last four years, the sensitivity of up and running ground-based detectors has been increasing \cite{Ab.al2.17,Ab.al.18,Ak.al.19}, and planned space-based missions such as LISA \cite{Am.al.17} will continue to bring new data, allowing the exploration of relativistic dynamics and fundamental physics in extreme astrophysical environments \cite{SaSc.09,Ba.al.20}. Such observations are made possible thanks to highly accurate template waveforms \cite{BuSa.15}, built by using a combination of numerical relativity simulations \cite{Sp.15,DuZl.18} and analytical approximation methods, such as post-Newtonian (PN) theory \cite{Bl.14}, effective field theory \cite{Po.16,Le.20}, black hole perturbation theory and the gravitational self-force (GSF) framework \cite{Po2.15,BaPo.18}, as well as effective one-body (EOB) models \cite{DaNa}.

In a seminal paper, Detweiler \cite{De.08} performed the first comparison of the predictions from the perturbative GSF framework and PN theory, for a binary system of nonspinning compact objects moving along a closed circular orbit. This comparison relied on the calculation of the so-called ``redshift parameter'' $U \equiv \ud t / \ud \tau$, i.e. on the time component of the 4-velocity of the smaller body, modelled as a structureless point particle, as a function of the angular velocity $\Omega$ of the binary's circular orbit. This comparison was later extended and refined in Refs.~\cite{Bl.al.10,Bl.al2.10,Bl.al.11,Bl.al.14,Bl.al2.14}.

Soon after, the discovery of the first law of compact binary mechanics \cite{Fr.al.02,Le.al.12,Bl.al.13} showed that the kinematical redshift parameter $U$ also has a \textit{dynamical} relevance, as it can be related in a simple manner to some of the global properties of the binary system, such as its binding energy and orbital angular momentum; see also \cite{Fu.al.17}. Using the first law of binary mechanics, the redshift parameter has been used to compare the predictions of GSF theory to the results from numerical relativity simulations of black hole binaries moving on quasi-circular orbits, showing remarkably good agreement, even for comparable-mass binary systems \cite{Le.al2.12,Zi.al.16,LeGr.18}.

Using the black hole perturbative techniques developed in \cite{Ma.al.96,MaTa.97}, the GSF contribution to the redshift parameter has been computed analytically, up to very high PN orders \cite{BiDa2.14,BiDa.15,Ka.al.15,Ka.al.16,Bi.al2.18,Bi.al.20}, and numerically with extremely high accuracy \cite{Sh.al.14,Jo.al.15}. Combined with the first law of binary mechanics, those results have been used to complete the calculation of the 4PN dynamics of arbitrary-mass-ratio binary systems of nonspinning compact objects \cite{Da.al.16}, as well as calibrate one of the potentials entering the EOB Hamiltonian that controls the conservative part of the binary's orbital dynamics \cite{Ba.al.12,Ak.al.12,BiDa.16}.

For compact binary systems moving along eccentric orbits, a generalization of the redshift parameter has been introduced in the context of GSF theory \cite{BaSa.11}, and then used to perform a comparison with the predictions from PN theory \cite{Ak.al.15}. This generalized redshift $\langle U \rangle$ appears naturally in the first law of mechanics for eccentric-orbit compact binaries \cite{Le.15,BlLe.17}, and it has been used to calibrate the remaining potentials that enter the conservative part of the EOB Hamiltonian \cite{Bi.al.16,AkvdM.16}.

Given the large amount of work, reviewed above, that relies on the redshift parameter $U$, it is arguably important to give this physical quantity a simple, geometrical and coordinate-invariant meaning. To the best of our knowledge, this question has only been adressed in the context of the perturbative GSF framework, where the redshift parameter $U$ has been shown to be gauge-invariant under gauge transformations generated by a helically symmetric gauge vector \cite{De.08,Sa.al.08,Sh.al.12}.

\subsection{Summary}

In this paper, we shall consider an arbitrary-mass-ratio binary system of spinning compact objects with internal structure, moving along a circular orbit with constant angular velocity $\Omega$. Mathematically, the approximation of an exactly closed circular orbit translates into the existence of a \textit{helical} Killing vector field $k^a$, so that $\Lik g_{ab} = 0$. The two compact bodies will be modelled by means of the gravitational skeleton formalism. Up to quadrupolar order, each of the two bodies is characterized by a 4-momentum $p^a$, a spin tensor $S^{ab}$ and a quadrupole tensor $J^{abcd}$, all defined along a worldline $\gamma$ with tangent 4-velocity $u^a$.

One of the main results established in this paper is that the worldline of each quadrupolar particle is an integral curve of the helical Killing field, namely
\beq\label{k=zu_intro}
    k^a|_\gamma = z u^a \, ,
\eeq
where $z$ is a constant along $\gamma$. \alext{The constants $z_1$ and $z_2$ of the two particles coincide only if the bodies share the same masses and spins \cite{De.08,LeGr.18}.} By using a spherical-type coordinate system $(t,r,\theta,\phi)$ adapted to the helical isometry, such that $k^a =  (\partial_t)^a + \Omega \, (\partial_\phi)^a$ in a neighborhood of $\gamma$, the coordinate components of the 4-velocity $u^a$ then simply read $u^\alpha = z^{-1} (1,0,0,\Omega)$. In particular,
\beq\label{z=1/U}
    z = \frac{\ud \tau}{\ud t} = \frac{1}{U} \, ,
\eeq
which shows that, while using adapted coordinates, Detweiler's redshift parameter is simply the inverse of the geometrically-defined constant that appears in the helical constraint \eqref{k=zu_intro}. The formulas \eqref{k=zu_intro} and \eqref{z=1/U} have been written down in various papers, e.g. in \cite{De.08,Ke.al2.10,Sh.al.12,Le2.14,Po.14,Ka.al.16}, but have never been proven in the more general context considered here, namely without any restriction on the coordinate system, for an \textit{arbitrary}-mass-ratio binary system of \textit{spinning} compact bodies \textit{with internal structure}. Another important result established in this paper is that the 4-velocity, the 4-momentum, the spin tensor and the quadrupole tensor of each particle are all Lie-dragged along $k^a$, and thus along $\gamma$ thanks to the formula \eqref{k=zu_intro}:
\beq\label{Lies}
    \Lik u^a = \Lik p^a = \Lik S^{ab} = \Lik J^{abcd} = 0 \, .
\eeq
Interestingly, results analogous to \eqref{k=zu_intro} and \eqref{Lies} have previously been established \cite{ScSt.81} in the framework of Dixon's theory of the dynamics of extended fluid bodies \cite{Di.79}, in a mathematical physics context. We chose instead to rely on the multipolar gravitational skeleton formalism which, together with a regularization method, is well-adapted to account for the self-field of the bodies. Indeed, this paper is the first in a series of articles that aims at extending the first law of binary mechanics \cite{Fr.al.02,Le.al.12,Bl.al.13} up to quadrupolar order, for spinning compact objects with internal structure moving along circular orbits. In particular, many of the results that will be established here will be used in \alext{future work} to give a geometrical derivation of the first law of compact binary mechanics at dipolar order, as a first step before accounting for the internal structure of the compact bodies. Our ability to derive the first law of mechanics crucially relies on the use of the gravitational skeleton formalism reviewed in the next section.

The remainder of this paper is organized as follows. The gravitational skeleton formalism that we shall use to model compact objects is reviewed in Sec.~\ref{sec:skeleton}. In Sec.~\ref{sec:colinear}, this formalism is applied to a binary system of quadrupolar particles moving along an exactly circular orbit, and the associated helical Killing field is proven to obey the constraint \eqref{k=zu_intro}. The multipole moments of the two particles are then shown to be Lie-dragged along the helical Killing field in Sec.~\ref{sec:Lips}. Finally, various conserved quantities associated with the helical isometry are explored in Sec.~\ref{sec:conserved}. A large amount of technical material is relegated to appendices. We first summarize our conventions and notations in App.~\ref{app:conventions}. We then review the concepts of bitensor, parallel propagator, invariant Dirac distribution and Lie derivative in App.~\ref{app:BiDiLie}. In App.~\ref{app:thm}, we recall two important theorems for distributional multipolar expansions, extending the second theorem to the case of an arbitrary number of multipolar particles. Using those results, the normal form of a quadrupolar gravitational skeleton is derived in App.~\ref{app:reduc}. Finally, we collect many well-known (and some less known) results on Killing vector fields in App.~\ref{app:Killing}.

\section{Gravitational skeleton}\label{sec:skeleton}

In this section, we review the multipolar, ``gravitational skeleton'' formalism that will be used throughout this paper to model spinning compact objects. This formalism dates back to 1937, with the pioneering works of Mathisson \cite{Ma.37,Ma.40} and Papapetrou \cite{Pa.51}, in an attempt to describe the motion of extended bodies in general relativity, going beyond the test particle approximation. This work was later made rigorous and systematic by Tulczyjew \cite{Tu.57,Tu.59}.

Related lines of work were developed during the 70's by Bailey and Israel \cite{BaIs.75}, who gave a Lagrangian formulation for the relativistic motion of extended bodies, and by Dixon \cite{Di.64,Di.73,Di.74} who introduced a multipole moment formalism for fluid bodies. More recently, Harte \cite{Ha.12,Ha.15} developed a formalism based on generalized Killing vectors, and extended Dixon's work by accounting for the self-field of the extended bodies.

\subsection{Newtonian and Einsteinian gravitational skeleton}\label{subsec:skel}

In the context of Newtonian gravitation, the mass density $\rho(t,\mathbf{x})$ of an arbitrary, extended body enclosed in a volume $\mathscr{B} \subset \mathbb{R}^3$ generates the Newtonian gravitational potential
\beq\label{U}
	U(t,\mathbf{x}) = \int_\mathscr{B} \frac{\rho(t,\mathbf{x}')}{\Vert \mathbf{x} - \mathbf{x}' \Vert} \, \ud^3 x' \, .
\eeq
Outside of that body, and given some reference point $\mathbf{x_\circ} \!\in\! \scB$, the gravitational potential \eqref{U} can be written as a multipolar expansion over the body's time-dependent mass multipole moments $I^{i_1\cdots i_\ell}(t) \equiv \int_\mathscr{B} \rho(t,\mathbf{x}) \, (x-x_\circ)^{i_1}\cdots (x-x_\circ)^{i_\ell} \, \ud^3 x$, according to (see e.g. Ref.~\cite{PoWi})
\beq\label{multiU}
	U(t,\mathbf{x}) = \sum_{\ell = 0}^\infty \frac{(-)^\ell}{\ell!} \, I^{i_1 \cdots i_\ell}(t) \, \partial_{i_1 \cdots i_\ell} {\Vert \mathbf{x}-\mathbf{x}_\circ \Vert}^{-1} \, .
\eeq

The idea of the gravitational skeleton model is to substitute the \textit{smooth} mass density $\rho$, whose support is the volume $\mathscr{B}$, by a \textit{distributional} mass density, say $\rho_\text{skel}$, whose support is a single point $\mathbf{x_\circ}\in\mathscr{B}$. It is given by a sum of time-dependent multipole moments, according to
\beq\label{rho_ske}
	\rho_\text{skel}(t,\mathbf{x}) = \sum_{\ell = 0}^\infty \frac{(-)^\ell}{\ell!} \, I^{i_1 \cdots i_\ell}(t) \, \partial_{i_1 \cdots i_\ell} \delta(\mathbf{x}-\mathbf{x}_\circ) \, ,
\eeq
where $\delta$ is the ordinary, 3-dimensional Dirac distribution. As $\nabla^2 \Vert \mathbf{x}-\mathbf{x}_\circ \Vert^{-1} = - 4\pi \, \delta(\mathbf{x}-\mathbf{x}_\circ)$, the multipolar potential \eqref{multiU} is a solution of the Poisson equation $\nabla^2 U = - 4\pi \, \rho_\text{skel}$, in the sense of distributions, for the singular mass density \eqref{rho_ske}. The mass density $\rho_\text{skel}$ can be seen as a \textit{particle} that is gravitationally equivalent to the original extended body, as it generates the same gravitational potential $U$.

\begin{figure}[t!]
    \begin{center}
    	\includegraphics[width=0.72\linewidth]{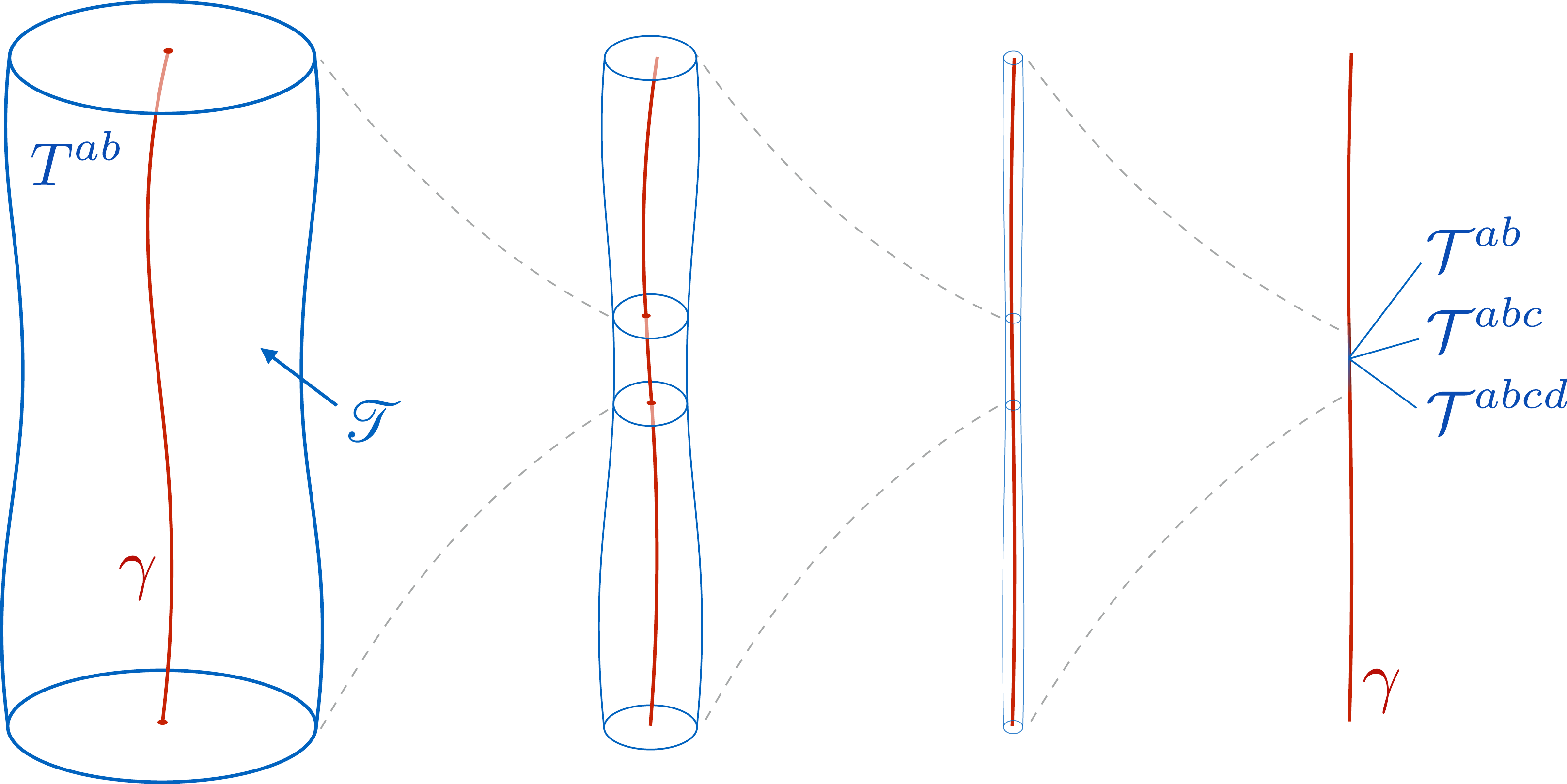}
        \caption{In the multipolar gravitational skeleton model, the smooth energy-momentum tensor $T^{ab}$ of an extended compact body enclosed in a worldtube $\mathscr{T}$ (left) is replaced by the distributional energy-momentum tensor \eqref{skelexp} of a particle endowed with a collection of multipoles $(\mathcal{T}^{ab},\mathcal{T}^{abc},\mathcal{T}^{abcd},\dots)$, all defined along a worldline $\gamma \subset \mathscr{T}$ (right), as if the extended body was ``observed from far away.''}
        \label{fig:shrink}
    \end{center}
\end{figure}

In the context of general relativity, by analogy with the Newtonian case, one then substitutes the smooth energy-momentum tensor $T^{ab}(x)$ of an arbitrary fluid ball, whose support is a worldtube $\mathscr{T} \subset \mathcal{M}$ in the spacetime manifold $\mathcal{M}$, by a distributional sum of multipole moments, say $T^{ab}_\text{skel}(x)$, whose support is restricted to a single worldline $\gamma \subset \mathscr{T}$, just as if the extended body was ``observed from far away" (see Fig.~\ref{fig:shrink}). The Ansatz for the gravitational skeleton is thus written as a covariant multipolar expansion of the type
\beq \label{skelexp}
	T^{ab}_\text{skel}(x) = \sum_{\ell = 0}^\infty \nabla_{c_1 \cdots c_\ell} \int_{\gamma} \mathcal{T}^{abc_1 \cdots c_\ell}(y) \,
\delta_4(x,y) \, \ud \tau \, ,
\eeq
where $\delta_4(x,y)$ is the invariant 4-dimensional Dirac distribution (defined in App.~\ref{app:BiDiLie}), with the four functions $y^\alpha(\tau)$ parameterizing the worldline $\gamma$ by the proper time $\tau$, so that $u^a \!\equiv\! \dot{y}^a(\tau)$ is the 4-velocity tangent to $\gamma$. We also introduced the convenient notation $\nabla_{c_1 \cdots c_\ell} \equiv \nabla_{c_1} \cdots \nabla_{c_\ell}$. The relativistic multipoles $\mathcal{T}^{abc_1 \cdots c_\ell}(y)$ are defined covariantly along the worldline $\gamma$ and are arbitrary at this stage. Physically, the distributional energy-momentum tensor \eqref{skelexp} may be interpreted as that of a multipolar point particle.

Most importantly, Geroch and Traschen \cite{GeTr.87} have proven that, because of the nonlinearity of the Einstein equation, the concept of a point particle does not make mathematical sense in general relativity; the closest thing to a point particle in general relativity being a black hole. However, in the context of approximation methods (e.g. post-Newtonian theory \cite{Bl.14}), the Einstein equation can be coupled to a distributional source such as \eqref{skelexp} in a meaningful manner, provided that a regularization scheme (e.g. dimensional regularization \cite{HoVe.72,BoGi.72}) is used to handle the divergent self-field of each particle. The metric can then be shown to obey the vacuum Einstein equation in a regularized sense (e.g. \cite{Bl.al.98}). Hence, in this paper, we will always implicitly assume that such a regularization scheme is being used while evaluating various tensor fields along the wordline of a multipolar point particle. All tensorial equations should thus be interpreted in a regularized sense.

\subsection{Tulczyjew's reduction at quadrupolar order}

The Ansatz \eqref{skelexp} involves an infinite number of degrees of freedom encoded in the infinite set $\{\mathcal{T}^{ab},\mathcal{T}^{abc},...\}$ of multipoles. However, for an extended body of mass $m$ and characteristic size $s$, the typical amplitude $\mathcal{T}_\ell$ of the $2^{\ell}$-pole $\mathcal{T}^{abc_1 \cdots c_\ell}$ scales as
\beq
	\mathcal{T}_\ell \sim m s^\ell \, .
\eeq
In the Newtonian limit, this is easily seen from the definition of the mass multipole moments $I^{i_1 \cdots i_\ell}(t)$. On the other hand, the covariant derivatives $\nabla_{c_1 \cdots c_\ell}$ typically scale as $1/r^\ell$, where $r$ is a coordinate measure of the distance to the center-of-mass of the actual extended body. Hence, \textit{qualitatively}, each additional multipole involves an extra factor of $s/r \sim m/r$ if the body is compact. As long as $r \gg m$, one might thus expect that the main physical features of a generic compact object are well captured by the first few terms of the multipolar expansion \eqref{skelexp}, just as in Newtonian gravity.

By truncating the gravitational skeleton at \textit{quadrupolar} order, i.e., by keeping only the first three terms in the expansion \eqref{skelexp}, we obtain the Ansatz for the energy-momentum tensor of a quadrupolar particle as
\beq \label{AnsatzSET}
	T^{ab} = \int_{\gamma} \mathcal{T}^{ab} \delta_4 \, \ud \tau + \nabla_c \int_{\gamma} \mathcal{T}^{abc} \delta_4 \, \ud \tau + \nabla_{cd} \int_\gamma \mathcal{T}^{abcd} \delta_4 \, \ud \tau \, ,
\eeq
where from now on we will omit the points $x \in \mathcal{M}$ and $y \in \gamma$, as well as the subscript ``skel'' for brevity. We emphasize that the worldline $\gamma$ and the multipoles $\mathcal{T}^{ab}$, $\mathcal{T}^{abc}$ and $\mathcal{T}^{abcd}$ are arbitrary, except for their symmetry with respect to the indices $a$ and $b$. In particular, these tensors encode $10+40+160 = 210$ degrees of freedom, for only 10 in $T^{ab}$. However, as was shown by Tulczyjew \cite{Tu.59}, the redundant degrees of freedom can be removed by a process of ``reduction'' of the energy-momentum tensor. The general strategy consists in working out constraints on the multipole moments that appear in the Ansatz \eqref{AnsatzSET}, as a consequence of the local conservation of energy and momentum,
\beq\label{DT=0}
	\nabla_a T^{ab} = 0 \, .
\eeq
As a result of this reduction, one obtains (i) the reduced form for $T^{ab}$ that does not contain any redundant degree of freedom, and (ii) equations of evolution for the ten genuine degrees of freedom that characterize the quadrupolar particle. At dipolar order, this reduction is a simple consequence of the results presented in App.~\ref{app:reduc}. The extension to quadrupolar order was performed in Ref.~\cite{StPu.10}.

\subsection{Energy-momentum tensor and equations of evolution} \label{subsec:emtee}

If the quadrupolar energy-momentum tensor is to verify the conservation law $\nabla_a T^{ab}=0$, then, following the reduction process introduced above and reviewed in App.~\ref{app:reduc}, there must exist a vector $p^a$, a rank-2 tensor $S^{ab}$ and a rank-4 tensor $J^{abcd}$, all defined along the worldline $\gamma$ with unit tangent $u^a$, such that the generic quadrupolar energy-momentum tensor \eqref{AnsatzSET} explicitly reads \cite{StPu.10}
\beq\label{SET}
    T^{ab} = \int_{\gamma} \bigl( u^{(a} p^{b)} + \tfrac{1}{3} R_{cde}^{\phantom{cde}(a}J^{b)cde} \bigr) \, \delta_4 \, \ud \tau + \nabla_c \int_{\gamma} u^{(a} S^{b)c} \, \delta_4 \, \ud \tau - \tfrac{2}{3} \nabla_{cd} \int_{\gamma} J^{c(ab)d} \, \delta_4 \, \ud \tau \, .
 \eeq
The tensor fields $p^a$, $S^{ab}$ and $J^{abcd}$ will be refered to as the momentum, spin and quadrupole of the particle, respectively. The spin $S^{ab}$ is antisymmetric and the quadrupole $J^{abcd}$ has the same algebraic symmetries as the Riemann tensor $R_{abcd}$. 

While performing the reduction of the generic energy-momentum tensor \eqref{AnsatzSET}, one obtains not only its reduced form \eqref{SET}, but also the equations governing the evolution of $p^a$ and $S^{ab}$ along $\gamma$,
\begin{subequations}\label{EE}
	\begin{align}
        \dot{p}^a & = \frac{1}{2} R_{bcd}^{\phantom{bcd}a} S^{bc} u^d - \frac{1}{6}J^{bcde}\nabla^a R_{bcde}\, , \label{EoM} \\
		\dot{S}^{ab} & = 2 p^{[a} u^{b]} + \frac{4}{3}R_{edc}^{\phantom{edc}[a}J^{b]cde}  \, , \label{EoP}
 	\end{align}
\end{subequations}
where the overdot stands for the covariant derivative with respect to the proper time $\tau$, i.e. $\dot{p}^a \equiv u^c \nabla_c p^a$. In the equation of motion \eqref{EoM}, the force that drives the evolution of the 4-momentum $p^a$ involves a spin coupling to curvature and a quadrupolar contribution. This force is spacelike but not necessarily orthogonal to the 4-velocity, since $\dot{p}^a u_a = - \tfrac{1}{6} J^{bcde} \dot{R}_{bcde}$. The equation of precession \eqref{EoP} shows that the torque that drives the evolution of the spin tensor $S^{ab}$ has two contributions as well. The first one vanishes if, and only if, $p^a$ is colinear to $u^a$, while the second one is of quadrupolar nature.

The energy-momentum tensor \eqref{SET} and the equations of evolution \eqref{EE} are recovered by all the formalisms that aim at modelling extended compact objects by means of multipolar expansions; see e.g. Refs.~\cite{BaIs.75,StPu.10,Bu.al.13,Ma.15}. However, when considering the coordinate components $(u^\alpha,p^\alpha,S^{\alpha\beta},J^{\alpha\beta\gamma\delta})$ of the multipoles as functions of the proper time $\tau$ along $\gamma$, the equations of evolution \eqref{EE} do not form a closed system of ordinary differential equations (for a given spacetime geometry). For example, there is no ``equation of motion'' for the quadrupole $J^{abcd}$. Depending on the approximation scheme used, the completeness of the differential system \eqref{EE} is only recovered under additional physical assumptions. In particular, one needs to (i) specify a physical model for the quadrupole $J^{abcd}$ in terms of the source variables $(u^a,p^a,S^{ab})$ and the geometry $(g_{ab},R_{abcd})$ along $\gamma$, and to (ii) impose a so-called ``spin supplementary condition,'' which in the present formalism takes the form of an algebraic equation involving the spin tensor $S^{ab}$. These two additional physical assumptions will be discussed below in Secs.~\ref{subsec:quad} and \ref{subsec:SSC}, respectively.

Next, from the variables $u^a$, $p^a$ and $S^{ab}$, we introduce three positive scalar fields along $\gamma$: the rest mass $m$, the dynamical mass $\mu$ and the spin amplitude $S$, which read
\beq\label{defnorms}
	m \equiv - p^a u_a \, , \quad \mu^2 \equiv - p^a p_a \quad \text{and} \quad S^2 \equiv \frac{1}{2} S^{ab} S_{ab} \, .
\eeq
In general, none of these scalar fields is conserved along $\gamma$, and the masses $m$ and $\mu$ need not coincide \cite{StPu.10}. However, as we shall prove later in Sec.~\ref{subsec:other_conserved}, in the case of a binary system of quadrupolar particles moving on a circular orbit, $m$, $\mu$ and $S$ are conserved for each particle. Finally, contracting Eq.~\eqref{EoP} with $u_b$ readily implies the momentum-velocity relationship
\beq \label{p=mu}
	p^a = m u^a - \dot{S}^{ab} u_b + \frac{4}{3} R_{edc}^{\phantom{edc}[a} J^{b]cde} u_b \, .
\eeq
The 4-momentum $p^a = p^a_\text{kin} + p^a_\text{hid}$ is the sum of the timelike kinematic momentum $p^a_\text{kin} \equiv mu^a$ and of the spacelike hidden momentum (see e.g. Ref.~\cite{Gr.al2.10}) $p^a_\text{hid} \equiv h^a_{\phantom{a}b} p^b$, such that $u_a p^a_\text{hid} = 0$, where we introduced the projector orthogonal to $u^a$,
\beq\label{h}
    h_{ab} \equiv g_{ab} + u_a u_b \, .
\eeq
Those equations summarize the gravitational skeleton model of compact bodies at quadrupolar order, and they will form the basis of much of the calculations in this series of papers. 

\subsection{Rotationally-induced and tidally-induced quadrupole}\label{subsec:quad}

Up to dipolar order, the dynamics of a spinning compact object is \textit{universal}, in the sense that that the equations of evolution \eqref{EE} with $J^{abcd} = 0$ can be used to model the motion and spin precession of both black holes and neutrons stars. However, this property of universality is lost at the next quadrupolar order, as the internal structure of the compact body appears through the quadrupole tensor $J^{abcd}$, whose proper time evolution is not driven by a dynamic equation analogous to \eqref{EE}. Rather, a physical model for $J^{abcd}$ has to be specified.

In this series of articles, we are interested in binary systems of spinning compact objects with \textit{internal structure}, in the sense that we account for some of the physical effects related to their extended nature. Our model considers two such effects, encoded into the quadrupole tensor $J^{abcd}$ of each body: a \textit{spin-induced} component coming from the body's proper rotation, and a \textit{tidally-induced} component coming from the gravitational influence of its companion.

On the one hand it has been shown \cite{Po.06,PoRo2.08,St.11,Bu.al.13,Ma.15,LeSt.15,LeSt2.15} that any spinning compact body has an $O(S^2)$ spin-induced quadrupole tensor
\beq\label{J_spin}
	J_\text{spin}^{abcd} = \frac{3\kappa_2}{m} \, u^{[a} S^{b]e} S_e^{\phantom{e}[c} u^{d]} \, ,
\eeq
where $\kappa_2$ is a dimensionless constant that measures the quadrupolar ``polarisability'' of the body induced by its proper rotation, and such that $\kappa_2 = 1$ for an isolated (i.e. Kerr) black hole \cite{Th.80,Po.98} and $\kappa_2 \sim 4-8$ for a neutron star \cite{LaPo.99,PaAp.12,St.15}, depending on the equation of state.

On the other hand, under the effect of their mutual gravitational interaction, both bodies also carry a tidally-induced quadrupole tensor, which for a \textit{non}-spinning compact object is given by \cite{GoRo2.06,DaNa.10,Bi.al.12,StPu.12}
\beq\label{J_tidal}
	J_\text{tidal}^{abcd} = 3\mu_2 \, u^{[a} E^{b][c} u^{d]} - 2\sigma_2 \, u^{[a} B^{b]f} \varepsilon^{cd}_{\phantom{cd}ef} u^e - 2\sigma_2 \, u^{[c} B^{d]f} \varepsilon^{ab}_{\phantom{ab}ef} u^e \, .
\eeq
Here, $\mu_2$ and $\sigma_2$ are dimensionful constants that measure the quadrupolar ``polarisability'' of each compact body, respectively induced by the gravito-electric and gravito-magnetic tidal fields of the companion,
\beq\label{E-B}
    E_{ab} \equiv R_{acbd} u^c u^d \quad \text{and} \quad B_{ab} \equiv \star R_{acbd} u^c u^d \, ,
\eeq
which are both symmetric and orthogonal to $u^a$, where $\star R_{abcd} \equiv \tfrac{1}{2} \varepsilon_{ab}^{\phantom{ab}ef} R_{efcd}$ is the self-dual of the Riemann tensor, $\varepsilon_{abcd}$ being the canonical volume form associated with the metric $g_{ab}$.

The so-called quadrupolar ``tidal Love numbers'' $\mu_2$ and $\sigma_2$ vanish for a nonspinning black hole \cite{DaNa2.09,BiPo.09,KoSm.12,Ch.al2.13,Gu.15,LeCa.20,Le.al.20,Ch.20}. For a non-spinning neutron star of areal radius $R$,
\beq\label{lambda2_sigma2}
    \mu_2 = \frac{2}{3} R^5 k_2 \quad \text{and} \quad \sigma_2 = \frac{1}{48} R^5 j_2 \, ,
\eeq
with $k_2 \sim 0.05-0.15$ and $|j_2| \lesssim 0.02$, depending on the equation of state \cite{BiPo.09,DaNa2.09}. The effect of the spin of the compact objects on the tidal Love numbers was explored in Refs.~\cite{LaPo.15,Pa.al.15,Pa.al2.15,La.17,LeCa.20,Le.al.20,Ch.20}. The quadrupole model \eqref{J_tidal} assumes that each compact object responds \textit{adiabatically} and \textit{linearly} to the tidal field induced by the orbiting companion. A more realistic model would include a dynamical response of the body to the applied tidal field (see e.g. Refs.~\cite{Ch.al2.13,St.al.16}) and would account for nonlinear tidal effects (see e.g. footnote 2 in Ref.~\cite{DaNa.10}).

\subsection{Spin supplementary condition} \label{subsec:SSC}

Consider for now an extended body, whose support is a worldtube $\mathscr{T}$, as well as a reference worldline $\gamma_\circ \subset \mathscr{T}$, with unit timelike tangent $v^a$. The body's momentum $p^a$ and spin $S^{ab}$ can be defined as surface integrals over the energy-momentum distribution, the later depending also on the choice of reference worldline $\gamma_\circ$ \cite{Di.79,Di.15}. The six degrees of freedom encoded in the spin tensor $S^{ab}$ can equivalently be encoded in two spacelike vectors $S^a$ and $D^a$, both orthogonal to $v^a$, according to
\beq\label{zip!}
	S^{ab} = \varepsilon^{abcd} v_c S_d + 2 D^{[a} v^{b]} \quad \Longleftrightarrow \quad
	\begin{cases}
		\, S^a \equiv - \frac{1}{2} \varepsilon^{abcd} v_b S_{cd} \, , \\
		D^a \equiv - S^{ab} v_b \, .
	\end{cases}
\eeq
Physically, the vector $D^a$ can be interpreted as the body's mass dipole moment, as measured by an observer with 4-velocity $v^a$, i.e., with respect to the reference worldline $\gamma_\circ$, while $S^a$ can be interpeted as the body's spin with respect to that wordline \cite{CoNa.15}. 

We now go back to the quadrupolar particle described by \eqref{SET}--\eqref{EE}. In a given basis, the equations of evolution \eqref{EE} are equivalent to a system of 10 \alext{ordinary} differential equations for 13 unknowns, namely the $4+6+3 = 13$ independent components of $p^{\alpha}(\tau)$, $S^{\alpha \beta}(\tau)$ and $u^{\alpha}(\tau)$. (Having specified a physical model for the quadrupole $J^{abcd}$, its components $J^{\alpha\beta\gamma\delta}(\tau)$ are known functions of $u^{\alpha}(\tau)$, $p^{\alpha}(\tau)$ and $S^{\alpha \beta}(\tau)$.) Since we did not specify the worldline $\gamma$ representing the body, it is not surprising that such under-determinacy should occur in the gravitational skeleton model. To obtain a well-posed problem, three additional constraints on the spin tensor, known as a spin supplementary condition (SSC), thus have to be imposed, equivalent to the choice of the reference wordline $\gamma_\circ$ with tangent $v^a$ for the actual extended body. These constraints take the form $S^{ab} f_b \!=\! 0$, where $f^a$ is a timelike vector. \alext{For instance, one might} adopt the so-called Frenkel-Mathisson-Pirani SSC
\beq\label{SSC}
	S^{ab} u_b = 0 \, ,
\eeq
which states that the mass dipole $D^a$ with respect to the wordline $\gamma$ vanishes, or equivalently that the spin vector $S^a$ in Eq.~\eqref{zip!} is orthogonal to the 4-velocity $u^a$. This natural choice of SSC is primarily motivated by the fact that the first law of compact binary mechanics to be derived in \alext{future work} will take its simplest form if \eqref{SSC} holds. Moreover, as we shall prove in the next Sec.~\ref{sec:colinear}, for circular orbits the 4-velocity of each particle is necessarily tangent to the generator $k^a$ of the helical Killing symmetry. For such orbits, the SSC \eqref{SSC} will thus be equivalent to the geometrically-motivated, algebraic constaint
\beq
	S^{ab} k_b = 0 \, .
\eeq
Other choices of SSC are of course possible, such as the Tulczujew-Dixon SSC $S^{ab} p_b = 0$. See for instance Refs.~\cite{KySe.07,CoNa.15,Co.al.18} for reviews of their well-posedness and physical interpretations.

\section{Quadrupolar particles and helical Killing isometry}\label{sec:colinear}

From now on, we shall consider a \textit{binary} system of spinning compact objects with internal structure, modelled in the multipolar gravitational skeleton framework that was introduced in the previous section, up to quadrupolar order. Except for the occurrence of a slow gradual inspiral driven by gravitational radiation-reaction, the orbits of (stellar mass) compact binaries can be considered to be \textit{circular}, to a good degree of approximation. Indeed, a post-Newtonian analysis shows that, to leading order in the radiation-reaction force, the rate of change of the angular velocity $\Omega$ of a compact binary system scales as $\dot{\Omega} / \Omega^2 \sim \nu \, (v/c)^5$ \cite{Bl.14}, where $\nu = m_1 m_2 / (m_1 + m_2)^2$ is the symmetric mass ratio and $v$ the typical orbital velocity. Consequently, during most of the inspiralling phase ($v/c \ll 1$) or for systems with large mass ratios ($\nu \ll 1$), we have $\dot{\Omega} / \Omega^2 \ll 1$, and the actual inspiral is well approximated by an adiabatic sequence of circular orbits.

Mathematically, the approximation of an exactly closed circular orbit
translates into the existence of a \textit{helical} Killing vector field $k^a$, along the orbits of which the spacetime geometry is invariant. Such a Killing field takes the form \cite{Fr.al.02,Go.al.02}
\beq\label{k}
    k^a = t^a + \Omega \, \phi^a \, ,
\eeq
where $t^a$ is timelike and $\phi^a$ is spacelike with integral curves of parameter length $2\pi$, while $\Omega$ is a constant that can be interpreted as the circular-orbit angular velocity of the binary. We emphasize that neither $t^a$ nor $\phi^a$ is a Killing field. The normalization of the helical Killing field \eqref{k} is chosen so that $k^a t_a \to -1$ at infinity.

As established in App.~\ref{app:Killing} (see in particular \eqref{LieRiem}--\eqref{LieSET} there), the energy-momentum tensor $T^{ab}$ must be invariant along the integral curves of any Killing field. Consequently, for a binary system of quadrupolar particles moving along an exactly closed circular orbit, we have
\beq\label{LieT}
    \mathcal{L}_k T^{ab} = 0 \, .
\eeq
Since the support of $T^{ab}$ is restricted to the two worldlines of the quadrupolar particles, say $\gamma_1$ and $\gamma_2$, we expect these worldlines to reflect the isometry generated by $k^a$. More precisely, for each particle $\ui \!\in\! \{1,2\}$, we expect $k^a$ to be colinear to the tangent 4-velocity $u^a_\ui$ along $\gamma_\ui$. We shall establish this central result in this section, \alext{Eq.~\eqref{k=zu} below}.

\alext{Importantly, we emphasize that the following calculations in Sec.~\ref{subsec:Lieconstraints} and \ref{subsec:quadKilling} hold for a \textit{generic} Killing vector field, and not merely for a helical Killing field of the form \eqref{k}. For the helically isometric case, however, the key relationship \eqref{k=zu} enforces the circular nature of the binary's orbital motion.}

\subsection{Lie-dragging constraints on the multipoles} \label{subsec:Lieconstraints}

To do so, it will be more convenient to use the generic form \eqref{AnsatzSET} of the energy-momentum tensor, rather than the reduced form \eqref{SET}, for each quadrupolar particle. Hence, our starting point is the energy-momentum tensor of a binary system of quadrupolar particles in the form
\beq \label{AnsatzSETbinary}
    T^{ab} = \sum_\ui \left\{ \int_{\gamma_\ui} \mathcal{T}^{ab}_\ui \, \delta^\ui_4 \, \ud \tau_\ui + \nabla_c \int_{\gamma_\ui} \mathcal{T}^{abc}_\ui \, \delta^\ui_4 \, \ud \tau_\ui + \nabla_{cd} \int_{\gamma_\ui} \mathcal{T}^{abcd}_\ui \, \delta^\ui_4 \, \ud \tau_\ui \right\} ,
\eeq
where we introduced the shorthands $\sum_\ui \!\equiv\! \sum_{\ui \in \{1,2\}}$ and $\delta_4^\ui \!\equiv\! \delta_4(x,y_\ui)$. More precisely, we shall use the \textit{normal form} associated with \eqref{AnsatzSETbinary}, which is obtained by performing an orthogonal decomposition of the multipoles $\mathcal{T}_\ui^{ab}$, $\mathcal{T}_\ui^{abc}$ and $\mathcal{T}_\ui^{abcd}$ with respect to the 4-velocity $u_\ui^a$. As shown in App.~\ref{app:reduc}, the energy-momentum tensor \eqref{AnsatzSETbinary} can be written in the alternative form
\beq\label{AnsatzSETnorm}
	T^{ab} = \sum_\ui \left\{ \int_{\gamma_\ui} \scT^{ab}_\ui \, \delta^\ui_4 \, \ud \tau_\ui + \nabla_c \int_{\gamma_\ui} \scT^{abc}_\ui \, \delta^\ui_4 \, \ud \tau_\ui + \nabla_{cd} \int_{\gamma_\ui} \scT^{abcd}_\ui \, \delta^\ui_4 \, \ud \tau_\ui \right\} ,
\eeq
where the dipole and quadrupole moments now obey the constraints $\scT_\ui^{abc} u^\ui_c=0$, $\scT_\ui^{ab[cd]}=0$ and $\scT_\ui^{abcd} u^\ui_d=0$. The fact that such a normal form always exists and is unique is one of the two theorems of Tuclzyjew, which are reviewed in App.~\ref{app:thm1} and \ref{app:thm2}. As shown in App.~\ref{app:reduc} (especially thanks to the formula \eqref{magic}), the multipole moments appearing in Eq.~\eqref{AnsatzSETnorm} are explicitly given by
\begin{subequations} \label{normalformquad}
	\begin{align}
		\scT_\ui^{ab} &\equiv \mathcal{T}_\ui^{ab} - \bigl( \mathcal{T}_\ui^{abu} - (\mathcal{T}_\ui^{abuu})\,\dot{} + 2 \mathcal{T}_\ui^{ab(cu)}\dot{u}^\ui_c \bigr)\, \dot{} \, + R_{cde}^{\phantom{cde}(a} \bigl( 2\mathcal{T}_\ui^{b)eu\hat{d}} u_\ui^c - \mathcal{T}_\ui^{b)e\hat{c}\hat{d}} \bigr) \, , \\
		\scT_\ui^{abc} &\equiv \mathcal{T}_\ui^{ab\hat{c}} - 2 \big( \mathcal{T}_\ui^{ab(du)} \bigr) \, \dot{} \, h^c_{\ui \; d} - \mathcal{T}_\ui^{abuu} \dot{u}_\ui^c \, , \\
		\scT_\ui^{abcd} &\equiv \mathcal{T}_\ui^{ab(\hat{c}\hat{d})} \, ,
	\end{align}
\end{subequations}
where, for each particle, the upper index $u$ denotes a contraction with $u^a$, e.g. $\mathcal{T}_\ui^{abu} \equiv \mathcal{T}_\ui^{abc} u_c$, and the hat above an index denotes a contraction with the orthogonal projector \eqref{h}, e.g. $\mathcal{T}_\ui^{ab\hat{c}} \equiv \mathcal{T}_\ui^{abd} h^c_{\phantom{c}d}$.

Heuristically, one expects that the Lie-dragging along $k^a$ of $T^{ab}$, Eq.~\eqref{LieT} above, implies some differential relationships obeyed by the multipoles $\scT_\ui^{ab}$, $\scT_\ui^{abc}$ and $\scT_\ui^{abcd}$ that appear in Eq.~\eqref{AnsatzSETnorm}. However, these multipoles are merely defined along $\gamma_\ui$. To define them as tensor fields off these worldlines, we introduce some smooth extensions $\tilde{\scT}_\ui^{ab}$, $\tilde{\scT}_\ui^{abc}$ and $\tilde{\scT}_\ui^{abcd}$. Such an extension can be chosen freely. Here, it is defined by parallel propagation along spacelike geodesics perpendicular to $\gamma_\ui$. Therefore, for each particle, in a normal neighborhood of a given point $y \in \gamma$, we define the extensions
\begin{subequations}\label{extensions}
	\begin{align}
		\tilde{\scT}^{ab}(x) &\equiv g^a_{\phantom{a}a'}(x,y)  \, g^b_{\phantom{b}b'}(x,y) \, \scT^{a'b'}(y) \, , \\
		\tilde{\scT}^{abc}(x) &\equiv g^a_{\phantom{a}a'}(x,y)  \, g^b_{\phantom{b}b'}(x,y)  \, g^c_{\phantom{c}c'}(x,y) \, \scT^{a'b'c'}(y) \, , \\
		\tilde{\scT}^{abcd}(x) &\equiv g^a_{\phantom{a}a'}(x,y)  \, g^b_{\phantom{b}b'}(x,y)  \, g^c_{\phantom{c}c'}(x,y) \,
		g^d_{\phantom{d}d'}(x,y) \,
		\scT^{a'b'c'd'}(y) \, ,
	\end{align}
\end{subequations}
where the bitensor $g^a_{\phantom{a}a'}(x,y)$ is the parallel propagator from $y$ to $x$ (see App.~\ref{subapp:paral}). As shall be proven in the next Sec.~\ref{subsec:quadKilling}, the final results will not depend upon this particular choice of extension. Owing to the presence of the invariant Dirac functional $\delta_4(x,y_\ui)$ in each integral in Eq.~\eqref{AnsatzSETnorm}, we may replace the multipoles by their smooth extensions \eqref{extensions} there. Taking the Lie derivative along $k^a$ on both sides and using Eq.~\eqref{LieT}, as well as the property \eqref{Liek-delta4} and the commutation \eqref{Commutation} of the Lie and covariant derivatives, we readily obtain
\beq \label{Li}
    \sum_\ui \left\{ \int_{\gamma_\ui} \scL^{ab}_\ui \, \delta^\ui_4 \, \ud \tau_\ui + \nabla_c \int_{\gamma_\ui} \scL^{abc}_\ui \, \delta^\ui_4 \, \ud \tau_\ui + \nabla_{cd} \int_{\gamma_\ui} \scL^{abcd}_\ui \, \delta^\ui_4 \, \ud \tau_\ui \right\} = 0 \, ,
\eeq
where we introduced the notation $\scL_\ui^{ab}\equiv \Lik \tilde{\scT}_\ui^{ab}$, $\scL_\ui^{abc}\equiv \Lik \tilde{\scT}_\ui^{abc}$ and $\scL_\ui^{abcd}\equiv \Lik \tilde{\scT}_\ui^{abcd}$ for the Lie derivatives of the smoothly extended multipoles.

The multipolar sums in Eq.~\eqref{Li} are not in normal form: the multipoles $\scL_\ui^{abc}$ and $\scL_\ui^{abcd}$ have the right algebraic symmetries, but they need not be orthogonal to $u^\ui_c$. However, thanks to the Thm.~\ref{thm1} in App.~\ref{app:thm1}, this normal form exists and is unique, and as shown in App.~\ref{app:reduc} it reads
\begin{align}\label{Linormal}
    \sum_\ui &\left\{ \int_{\gamma_\ui} \Bigl[ \scL^{ab}_\ui - \bigl( \scL_\ui^{abu} + \scL_\ui^{abcu}\dot{u}^\ui_c - (\scL_\ui^{abcu})\,\dot{}\,u^\ui_c \bigr)\,\dot{} \, + 2 R_{cde}^{\phantom{cde}(a} \scL_\ui^{b)eu\hat{d}} u_\ui^c \Bigr] \, \delta^\ui_4 \, \ud \tau_\ui \right. \nonumber \\ &\left. \; + \, \nabla_c \int_{\gamma_\ui} \bigl[ \scL_\ui^{ab\hat{c}} - 2 (\scL_\ui^{abdu}) \, \dot{} \, h^c_{\ui \; d} - \scL_\ui^{abuu} \dot{u}_\ui^c \bigr] \, \delta^\ui_4 \, \ud \tau_\ui + \nabla_{cd} \int_{\gamma_\ui} \scL_\ui^{ab\hat{c}\hat{d}} \, \delta^\ui_4 \, \ud \tau_\ui  \right\} = 0 \, .
\end{align}
Those multipolar sums are now in normal form, so that, according to Thm.~\ref{thm2} in App.~\ref{app:thm2}, each integrand must be identically equal to zero along $\gamma_\ui$. This implies the constraints
\begin{subequations}\label{constraints}
    \begin{align}
        \scL_\ui^{ab} &= \bigl( \scL_\ui^{abu} + \scL_\ui^{abcu}\dot{u}^\ui_c - (\scL_\ui^{abcu})\,\dot{}\,u^\ui_c \bigr)\,\dot{} \, - 2 R_{cde}^{\phantom{cde}(a}
        \scL_\ui^{b)eu\hat{d}} u_\ui^c  \, , \\
        \scL_\ui^{abc} &= - \scL_\ui^{abu} u_\ui^c + 2 (\scL_\ui^{abdu}) \, \dot{} \, h^c_{\ui \; d} + \scL_\ui^{abuu} \dot{u}_\ui^c \, , \\
        \scL_\ui^{abcd} &= - \scL_\ui^{abuu} u_\ui^c u_\ui^d + 2\scL_\ui^{abu(\hat{d}} u_\ui^{c)} \, .
    \end{align}
\end{subequations}
These constraints will be central to prove, in the next subsection, that quadrupolar particles do follow helical Killing trajectories.

\subsection{Quadrupolar particles follow Killing trajectories} \label{subsec:quadKilling}

Let $f_{ab}$ denote a tensor field with compact support $\scV\subset\mathcal{M}$ that is smooth on the interior $\scV^\circ$ of $\scV$. The Lie dragging \eqref{LieT} of the distributional energy-momentum tensor \eqref{AnsatzSETnorm} implies that $\int_\scV f_{ab} \Lik T^{ab} \, \ud V = 0$, where $\ud V$ is the invariant 4-volume element. Therefore, by using the Leibniz rule on the Lie derivative, we readily obtain
\beq\label{Adam}
	\int_\scV \Lik(T^{ab} \! f_{ab}) \, \ud V = \int_\scV T^{ab} \Lik f_{ab} \, \ud V \, .
\eeq
The integral appearing in the left-hand side of Eq.~\eqref{Adam} is easily shown to vanish, as follows. By using the definition of the Lie derivative of a scalar field, together with $\nabla_c k^c = 0$, and applying Stokes' theorem together with $f_{ab} = 0$ on the boundary $\partial \scV$, we have
\beq\label{tempa}
	\int_\scV \Lik(T^{ab} \! f_{ab}) \, \ud V = \int_\scV \nabla_c (k^c T^{ab} \! f_{ab}) \, \ud V = \oint_{\partial \scV} T^{ab} \! f_{ab} \, k^c \ud \Sigma_c = 0 \, ,
\eeq
where $\ud\Sigma_c$ is the surface element orthogonal to $\partial\scV$. Next, we substitute the expression \eqref{AnsatzSETnorm} of the binary's quadrupolar energy-momentum tensor, in normal form, into the integral that appears in the right-hand side of Eq.~\eqref{Adam}. After commuting the integrals over $\scV$ and $\gamma_\ui$, integrating by parts, using Stokes' theorem and the compact-supported nature of the tensor $f_{ab}$, as well as the defining property \eqref{delta4} of the invariant Dirac distribution, we obtain
\beq\label{tempo}
    \int_\scV T^{ab} \Lik f_{ab} \, \ud V = \sum_\ui \int_{\gamma_\ui} \big( \scT^{ab}_\ui \Lik f_{ab}  - \scT^{abc}_\ui \nabla_c \Lik f_{ab} + \scT^{abcd}_\ui \nabla_{cd} \Lik f_{ab}	\bigr) \, \ud \tau_\ui \, .
\eeq
On the one hand, from the result \eqref{Commutation} we may commute the covariant and Lie derivatives in the second and third terms in the right-hand side of \eqref{tempo}. On the other hand, we notice that $\scT_\ui^{ab} = \tilde{\scT}_\ui^{ab}$ along $\gamma_\ui$, and similarly for $\scT_\ui^{abc}$ and $\scT_\ui^{abcd}$, so that the multipoles can be replaced by their smooth extensions \eqref{extensions} off $\gamma_\ui$. Combined with Eqs.~\eqref{Adam} and \eqref{tempa}, the formula \eqref{tempo} then implies
\beq\label{tempi}
    \sum_\ui \int_{\gamma_\ui} \big( \tilde{\scT}^{ab}_\ui \Lik f_{ab} - \tilde{\scT}^{abc}_\ui \Lik \nabla_c f_{ab} + \tilde{\scT}^{abcd}_\ui \Lik \nabla_{cd} f_{ab}	\bigr) \, \ud \tau_\ui = 0 \, .
\eeq
Applying the Leibniz rule to the Lie derivatives in the integrand and recalling the notation $\scL\equiv\Lik \tilde{\scT}$, the formula \eqref{tempi} now implies
\begin{align}\label{tempu}
    &\sum_\ui \int_{\gamma_\ui} \Lik \bigl( \tilde{\scT}_\ui^{ab} f_{ab} - \tilde{\scT}_\ui^{abc} \nabla_cf_{ab} + \tilde{\scT}_\ui^{abcd} \nabla_{cd} f_{ab} \bigr) \, \ud \tau_\ui \nonumber \\
    =& \sum_\ui \int_{\gamma_\ui} \bigl( \scL_\ui^{ab} f_{ab} - \scL_\ui^{abc} \nabla_cf_{ab} + \scL_\ui^{abcd} \nabla_{cd} f_{ab} \bigr) \, \ud \tau_\ui \nonumber \\
    =& \sum_\ui \int_{\gamma_\ui} \left[ \scL_\ui^{abu} f_{ab} - (\scL_\ui^{abcd}f_{ab}) \, \dot{} \, u^\ui_c u^\ui_d - 2\scL_\ui^{abuc}\nabla_c f_{ab} \; \right] \dot{} \; \ud \tau_\ui = 0 \, ,
\end{align}
where we have used the constraints \eqref{constraints} in the second equality and the fact that $f_{ab}$ has a compact support in the last equality. Equation \eqref{tempu} must hold for any $f_{ab}$ with compact support $\scV$ and smooth on $\scV^\circ$. In particular, it must hold for all tensor fields $f_{ab}$ whose support excludes either $\gamma_1$ or $\gamma_2$, such that both proper time integrals in \eqref{tempu} must identically vanish. Therefore, for all $\ui \in \{1,2\}$, we have established that
\beq\label{tempy}
	\int_{\gamma_\ui} \Lik f_\ui \, \ud \tau_\ui = 0 \, , \quad \text{where} \quad f_\ui \equiv \tilde{\scT}_\ui^{ab} f_{ab} - \tilde{\scT}_\ui^{abc} \nabla_cf_{ab} + \tilde{\scT}_\ui^{abcd} \nabla_{cd} f_{ab} \, .
\eeq
Clearly, having $k^a \propto u_\ui^a$ along $\gamma_\ui$ is a \textit{sufficient} condition for Eq.~\eqref{tempy} to hold for any $f_{ab}$. Indeed, $k^a \!\propto\! u_\ui^a$ implies $\Lik f_\ui = k^a \nabla_a f_\ui \propto \dot{f}_\ui$, and the integral of $\dot{f}_\ui(\tau_\ui)$ over $\gamma_\ui$ vanishes because $f_\ui$ has compact support. We now argue that $k^a\propto u_\ui^a$ along $\gamma_\ui$ is also a \textit{necessary} condition for Eq.~\eqref{tempy} to hold for all $f_{ab}$. 

We summarize here the idea behind the proof and relegate the details to App.~\ref{subapp:detailsfab}. First, we perform an orthogonal decomposition of $k^a$ with respect to the tangent 4-velocity $u_\ui^a$ to $\gamma_\ui$, according to $k^a|_{\gamma_\ui} = z_\ui u_\ui^a + w_\ui^a$, where $z_\ui \equiv - k^a u^\ui_a$ and $w_\ui^a \equiv h^a_{\ui \; b} k^b$. With these notations, the integrand in Eq.~\eqref{tempy} becomes $\Lik f_\ui=z_\ui\dot{f}_\ui + w^a_\ui \nabla_a f_\ui$. Second, we let $\FF$ denote the set of scalar fields $f_\ui$ given by Eq.~\eqref{tempy}, with $f_{ab}$ of compact support $\scV$ and smooth on $\scV^\circ$. We now consider the following proposition:
\beq\label{claim}
    \forall f_\ui \in \FF \, , \, \, \int_{\gamma_\ui} (z_\ui \dot{f}_\ui + w^a_\ui\nabla_a f_\ui) \, \ud \tau_\ui = 0 \quad \Longrightarrow \quad \forall y \in \gamma_\ui \, , \, \,
    \begin{cases}
        \dot{z}_\ui(y) = 0 \, , \\	 
        w^a_\ui(y) = 0 \, .
     \end{cases}
\eeq
Proposition \eqref{claim} is most easily proved by contraposition. More precisely, one assumes that $\dot{z}_\ui\neq 0$ or $w^a_\ui\neq 0$ at some point along $\gamma_\ui$ and shows that, consequently, there exists an $f_\ui\in\FF$ such that the integral on the left-hand side is nonzero; see App.~\ref{subapp:detailsfab} for details. Since this result holds for any 4-volume $\scV$ chosen initially, we conclude that $\dot{z}_\ui=0$ and $w^a_\ui = 0$ at any point along $\gamma_\ui$. Consequently, the expansion of $k^a$ along $\gamma_\ui$ simply reads $k^a|_{\gamma_\ui} = z_\ui u_\ui^a$, with $z_\ui$ constant along $\gamma_\ui$. This is one of the most important results in this paper.

\subsection{Detweiler redshift parameter}\label{subsec:Detweiler}

In the last paragraphs we have proven that if the energy-momentum tensor $T^{ab}$ describes a pair of quadrupolar particles moving along a circular orbit, then its Lie-dragging $\Lik T^{ab}=0$ along the helical Killing field $k^a$ implies that for any particle $\ui\in \{1,2\}$ of the system, there exists a constant scalar field $z_\ui$ defined on $\gamma_\ui$ such that
\beq\label{k=zu}
    \forall y \in \gamma_\ui \, , \quad k^a(y)=z_\ui u^a_\ui(y) \, .
\eeq
In other words, $k^a$ is tangent to the worldlines of both particles, or equivalently $\gamma_1$ and $\gamma_2$ are integral curves of the helical Killing field $k^a$. Moreover, since $u^a_\ui$ and $k^a|_{\gamma_\ui}$ are both timelike and future-directed, the constant $z_\ui$ is strictly positive along $\gamma_\ui$. That the helical Killing field \eqref{k} should be colinear to the particles' 4-velocities makes perfect physical sense, because the support of the helically symmetric energy-momentum tensor \eqref{AnsatzSETbinary} is restricted to the worldlines $\gamma_1$ and $\gamma_2$.

Following Detweiler's seminal work \cite{De.08}, the scalar field $z_\ui$ has been coinced the ``redshift'' parameter/variable, e.g. in Refs.~\cite{Bl.al.10,Ak.al.12,Po.14,Zi.al.16,LeGr.18}. Contracting \eqref{k=zu} with $u^\ui_a$ and taking the norm of \eqref{k=zu} yields two simple expressions for the redshift:
\beq\label{z}
	z_\ui = - u_\ui^a k_a \quad \text{and} \quad z_\ui = |k|_\ui \equiv (- k^a k_a)^{1/2}|_{\gamma_\ui} \, .
\eeq
In particular, the redshift coincides with the norm of $k^a$ along $\gamma_\ui$. Since the norm of a Killing field is necessarily conserved along its integral curves (see App.~\ref{app:Killing}), the redshift $z_\ui$ must be conserved along $\gamma_\ui$. Indeed, the constraint \eqref{k=zu} and Killing's equation \eqref{Killing} imply
\beq\label{zdot}
	z_\ui \dot{z}_\ui = - \tfrac{1}{2} \, u_\ui^a \nabla_a (k^b k_b) = - u_\ui^a k^b \nabla_a k_b = - z_\ui u_\ui^a u_\ui^b \nabla_{(a} k_{b)} = 0 \, .
\eeq
This is consistent with the result \eqref{claim}. The conserved redshift \eqref{z} is not to be confused with the Killing energy of the quadrupolar particle, which will be defined in Sec.~\ref{subsec:Killing_energy} below.

We stress that \eqref{k=zu} holds irrespective of a choice of SSC for the spins $S_\ui^{ab}$ of the particles, and irrespective of a particular physical model for the quadrupoles $J_\ui^{abcd}$. Moreover, while we have established this result at the quadrupolar order, we expect it to hold at \textit{any} order in the multipolar expansion \eqref{skelexp}. In particular, at the monopolar order it is a classical result that the solutions to the equations of motion \eqref{EoM} for nonspinning massive particles are timelike geodesics. Equation \eqref{k=zu} thus implies that the integral curves of the helical Killing vector along $\gamma_\ui$ must be timelike geodesics in this case.

Finally, we note that the constraint \eqref{k=zu} implies that for any scalar field $f$ defined along $\gamma_\ui$, the Lie derivative along $k^a$ simply reduces, up to a factor of the constant redshift \eqref{z}, to the ordinary derivative with respect to proper time $\tau_\ui$ along $\gamma_\ui$, namely
\beq\label{blibli}
	\Lik f |_{\gamma_\ui} = z_\ui \dot{f} = z_\ui \, \frac{\ud f}{\ud \tau_\ui} \, .
\eeq
Introducing a spherical-type coordinate system $(t,r,\theta,\phi)$ adapted to the helical Killing symmetry, such that $k^a =  (\partial_t)^a + \Omega \, (\partial_\phi)^a$ holds everywhere (or at least in a neighborhood of $\gamma_\ui$), the coordinate components of the 4-velocity $u_\ui^a$ simply read $u_\ui^\alpha \!=\! z_\ui^{-1} (1,0,0,\Omega)$. In particular, $z_\ui = \ud \tau_\ui / \ud t$ such that Eq.~\eqref{blibli} reduces to
\beq\label{blabla}
	\Lik f |_{\gamma_\ui} = \frac{\ud f}{\ud t} \, .
\eeq

\section{Lie-dragging of velocity, momentum, spin and quadrupole} \label{sec:Lips}

Thanks to the colinearity \eqref{k=zu} of the helical Killing field \eqref{k} and the tangent 4-velocity to the worldline $\gamma_\ui$, the Lie derivative along $k^a$ of any tensor field defined solely along $\gamma_\ui$ is well defined. In particular, $\Lik u_\ui^a$, $\Lik p_\ui^a$, $\Lik S_\ui^{ab}$ and $\Lik J_\ui^{abcd}$ are well-defined tensor fields along $\gamma_\ui$. In this section, we shall omit the subscript $\ui \in \{1,2\}$ whenever an equation applies for both particles. We shall establish that the 4-velocity $u^a$, momentum $p^a$, spin $S^{ab}$ and quadrupole $J^{abcd}$ of each particle are Lie-dragged along the helical Killing field $k^a$, as expected given the Lie-dragging \eqref{LieT} of the energy-momentum tensor of the binary system. 

\subsection{Lie-dragging of velocity and related identities}

Taking the covariant derivative of Eq.~\eqref{k=zu} along $\gamma$ readily gives $\dot{k}^a = z \dot{u}^a$, because the redshift $z$ is constant. Using Eq.~\eqref{k=zu}, this equation can be rewritten as $u^b \nabla_b k^a = k^b \nabla_b u^a$, which is equivalent to
\beq\label{Liu}
     \Lik u^a = 0 \, .
\eeq
Therefore, the 4-velocity $u^a$ is Lie-dragged along $k^a|_\gamma = z u^a$. Together with the Lie-dragging $\Lik g_{ab} = 0$ of the metric, the formula \eqref{Liu} implies that the projector $h_{ab} = g_{ab} + u_a u_b$ is also Lie-dragged along $k^a$, namely
\beq\label{Ligamma}
    \Lik h_{ab} = 0 \, .
\eeq
Moreover, for any tensor field $T^{a_1 \cdots a_k}_{\phantom{a_1 \cdots a_k} \; b_1 \cdots b_l}$ of type $(k,l)$, we may combine Eq.~\eqref{Liu} with the commutation \eqref{Commutation} of the Lie and covariant derivatives, together with the Leibnitz rule, to establish that
\beq\label{comLidot}
    \Lik \dot{T}^{a_1 \cdots a_k}_{\phantom{a_1 \cdots a_k} \; b_1 \cdots b_l} = \bigl( \Lik T^{a_1 \cdots a_k}_{\phantom{a_1 \cdots a_k} \; b_1 \cdots b_l} \bigr) \, \dot{} \; ,
\eeq
i.e., the Lie derivative along $k^a$ commutes with the covariant derivative along $\gamma$. This general result will prove useful in Sec.~\ref{subsec:other_conserved} below.

\subsection{Lie-dragging of momentum, spin and quadrupole}\label{subsec:Lies}

Our next objective is to establish that, for each particle, the momentum $p^a$, spin $S^{ab}$ and quadrupole $J^{abcd}$ are Lie-dragged as well. We shall introduce the following notations for the Lie derivatives along $k^a$ of these multipoles:
\beq \label{greeek}
    \pi^{a} \equiv \Lik p^a \, , \quad \Sigma^{ab} \equiv \Lik S^{ab} \, \quad \text{and} \quad \Theta^{abcd} \equiv \Lik J^{abcd} \, .
\eeq
Our starting point is the reduced form \eqref{SET} of the energy-momentum tensor of a quadrupolar particle. Using the Lie-dragging \eqref{LieT} of the energy-momentum tensor of a binary system of quadrupolar particles moving on a circular orbit, we obtain 
\begin{align}\label{quadred}
    0 = \sum_\ui &\left\{ \int_{\gamma_\ui} \Bigl[ u_\ui^{(a}\pi_\ui^{b)} + \tfrac{1}{3} R_{cde}^{\phantom{cde}(a}\Theta_\ui^{b)cde} \Bigr] \, \delta^\ui_4 \, \ud \tau_\ui \right. \nonumber \\ &\left. \; + \, \nabla_c \int_{\gamma_\ui} u_\ui^{(a}\Sigma_\ui^{b)c} \, \delta^\ui_4 \, \ud \tau_\ui - \tfrac{2}{3} \nabla_{cd} \int_{\gamma_\ui} \Theta_\ui^{c(ab)d} \, \delta^\ui_4 \, \ud \tau_\ui \right\} ,
\end{align}
where we have used the Lie-dragging \eqref{Liu}, \eqref{Liek-delta4} and \eqref{LieRiem} of the velocities $u_\ui^a$, the invariant Dirac functional $\delta_4(x,y_\ui)$ and the Riemann tensor $R_{abc}^{\phantom{abc}d}$, as well as the commutation \eqref{Commutation} of the Lie and covariant derivatives. Next, we may bring Eq.~\eqref{quadred} into its normal form, just like the energy-momentum tensor \eqref{AnsatzSETbinary} was brought into its normal form \eqref{AnsatzSETnorm}--\eqref{normalformquad}. From Tulczyjew's second theorem (see App.~\ref{app:thm2}), we then obtain constraints on the Lie-dragged multipoles $\pi_\ui^a$, $\Sigma_\ui^{ab}$ and $\Theta_\ui^{abcd}$, which read for each particle 
\begin{subequations}\label{constraints_pS}
	\begin{align}
        u^{(a}\pi^{b)} &= \bigl( u^{(a}\Sigma^{b)u} \bigr) \, \dot{} -\tfrac{1}{3} R_{cde}^{\phantom{cde}(a} \Theta^{b)cde} + \tfrac{2}{3} \bigl( \Theta^{u(ab)u} \bigr)\, \ddot{} + \tfrac{4}{3} R_{cde}^{\phantom{cde}(a} \Theta^{b)(\hat{d}u)e} u^c \, , \label{pS1} \\
		u^{(a}\Sigma^{b)\hat{c}} &= - \tfrac{4}{3} \bigl( \Theta^{d(ab)u} \bigr) \, \dot{} \, h^c_{\phantom{c}d} - \tfrac{2}{3} \Theta^{u(ab)u}\dot{u}^c \, , \label{pS2} \\
		\Theta^{\hat{c}(ab)\hat{d}} &= 0 \label{pS3} \, .
	\end{align}
\end{subequations}
These equations are the reduced form of the more general Eqs.~\eqref{constraints}. Let us now establish that Eqs.~\eqref{constraints_pS} imply the vanishing of the Lie-dragged multipoles \eqref{greeek}.

We start by showing that \eqref{pS3} implies $\Theta^{abcd}=0$. To this end, we perform an orthogonal decomposition of $\Theta^{abcd}$ with respect to the 4-velocity $u^a$, with help of the orthogonal projector \eqref{h}. Thanks to the algebraic symmetries of $\Theta^{abcd}$, namely those of the Riemann curvature tensor, this decomposition simply reads
\beq \label{decompTheta}
    \Theta^{abcd} = \hat{\Theta}^{abcd} + 2 u^{[a}\Phi^{b]cd} + 2 u^{[d}\Phi^{c]ba} - 4 u^{[a}\Psi^{b][c}u^{d]} \, ,
\eeq
where the tensors $\hat{\Theta}^{abcd} \equiv \Theta^{\hat{a}\hat{b}\hat{c}\hat{d}}$, $\Phi^{abc} \equiv \Theta^{\hat{a}u\hat{b}\hat{c}}$ and $\Psi^{ab} \equiv \Theta^{\hat{a}u\hat{b}u}$ are all orthogonal to $u^a$. We then symmetrize \eqref{decompTheta} with respect to the indices $b$ and $c$, and contract it with the projector $h^e_{\phantom{e}a} h^f_{\phantom{f}d}$, so that Eq.~\eqref{pS3} implies
\beq \label{decompTheta_3}
    \hat{\Theta}^{c(ab)d} + 2 \Phi^{(cd)(a}u^{b)} - u^a u^b \Psi^{cd} = 0 \, .
\eeq
Contracting \eqref{decompTheta_3} with $u_au_b$ and $u_ah^e_{\phantom{e}b}$  gives $\Psi^{ab}=0$ and $\Phi^{abc}=0$, respectively. Substituting these equations back into \eqref{decompTheta_3} yields the third constraint $\hat{\Theta}^{a(bc)d}=0$. Finally, substituting these three constraints into the decomposition \eqref{decompTheta} gives 
\beq \label{decompTheta_4}
    \Theta^{abcd} = \hat{\Theta}^{a[bc]d} \, ,
\eeq
which implies $\Theta^{a(bc)d} = 0$. Combined with the other algebraic symmetries of $\Theta^{abcd}$, namely those of the Riemann curvature tensor, this readily implies that $\Theta^{abcd}$ vanishes identically. We have thus proven that 
\beq \label{LieQuad}
\Theta^{abcd} \equiv \Lik J^{abcd} = 0 \, ,
\eeq
so that the quadrupole of each particle is Lie-dragged along its worldline.

Given Eq.~\eqref{LieQuad}, the system \eqref{constraints_pS} simplifies drastically, as it reduces to that for a dipolar particle, namely
\begin{subequations}\label{constraints_pS_simp}
	\begin{align}
        u^{(a}\pi^{b)} &= \bigl( u^{(a}\Sigma^{b)u} \bigr) \, \dot{} \, , \label{pS1_bis} \\
		u^{(a}\Sigma^{b)\hat{c}} &= 0 \, . \label{pS2_bis}
	\end{align}
\end{subequations}
Contracting Eq.~\eqref{pS2_bis} with $h^d_{\phantom{d}a} u_b$ and $u_a u_b$ implies $\Sigma^{ab} h^c_{\phantom{c}a} h^d_{\phantom{d}b}=0$ and $\Sigma^{ab}h^c_{\phantom{c}a} u_b=0$, respectively. Since $\Sigma^{ab} u_a u_b = 0$ by the antisymmetry of $\Sigma^{ab}$, we conclude that all the contributions to the orthogonal decomposition of $\Sigma^{ab}$ with respect to $u^a$ vanish. Consequently, we have shown that 
\beq \label{LieSpin}
\Sigma^{ab} \equiv \Lik S^{ab} = 0 \, ,
\eeq
so that the spin of each particle is Lie-dragged along its worldline. Finally, we may substitute Eq.~\eqref{LieSpin} into \eqref{pS1_bis} and contract with $h^c_{\phantom{c}a} u_b$ and $u_a u_b$ to obtain $\pi^a h^b_{\phantom{c}a}=0$ and $\pi^{a}u_a=0$. We thus conclude that
\beq \label{LieMom}
\pi^a \equiv \Lik p^{a} = 0 \, ,
\eeq
so that the 4-momentum of each particle is Lie-dragged along its worldline. Notice that the physical models \eqref{J_spin} and \eqref{J_tidal} for a rotationally-induced or tidally-induced quadrupole are consistent with the Lie-dragging of $u^a$, $p^a$, $S^{ab}$ and $J^{abcd}$. We naturally expect that the results \eqref{LieQuad}, \eqref{LieSpin} and \eqref{LieMom} extend to an arbitrary multipolar order in the gravitational skeleton formalism that was reviewed in Sec.~\ref{sec:skeleton}. Finally, together with Eq.~\eqref{Lie-epsilon}, the Lie-dragging \eqref{Liu} and \eqref{LieSpin} of the 4-velocity and spin tensor readily imply the Lie-dragging of the spin vector and mass dipole moment \eqref{zip!}:
\beq
    \Lik S^a = 0 \quad \text{and} \quad \Lik D^a = 0 \, .
\eeq

Interestingly, the formula \eqref{k=zu} and some of its consequences---namely the Lie dragging \eqref{LieSpin} and \eqref{LieMom} of the 4-momentum  and spin tensor---were previously established in \cite{ScSt.81}, in the context of Dixon's formalism for extended fluid bodies. However, this formalism is not well adapted to our purposes, namely to extend the first law of binary mechanics up to quadrupolar order. The consistency of our results with those of Ref.~\cite{ScSt.81} illustrates, once more, that these two formalisms are complementary and consistent with each other.

\subsection{Algebraic constraints on the multipoles}

Finally, we discuss an interesting consequence of the Lie-dragging \eqref{LieMom} and \eqref{LieSpin} of the momentum $p^a$ and spin $S^{ab}$, in light of the helical constraint \eqref{k=zu}. Combining Eqs.~\eqref{k=zu} and \eqref{Killing}, the formulas \eqref{LieMom} and \eqref{LieSpin} can be rewritten as
\begin{subequations}
	\begin{align}
		z \dot{p}_a &= p^c \nabla_c k_a = - p^c \nabla_a k_c \, , \label{zdotp} \\
		z \dot{S}_{ab} &= 2 S^c_{\phantom{c}[a} \nabla_{b]} k_c = 2 \nabla_c k_{[a} S^c_{\phantom{c}b]} \label{zdotS} \, .
	\end{align}
\end{subequations}
By combining those Lie-dragging equations with the equations of evolution \eqref{EE}, while using the helical constraint \eqref{k=zu}, we obtain the following relations that must be satisfied by the momentum, spin and quadrupole of each particle:
\begin{subequations}\label{algebraic}
	\begin{align}
        p^c \nabla_c k_a &= \frac{1}{2} R_{bcda} S^{bc} k^d - \frac{1}{6} \, z J^{bcde} \nabla_a R_{bcde} \, , \label{bibop} \\
 		S^c_{\phantom{c}[a} \nabla_{b]} k_c &= p_{[a} k_{b]} + \frac{2}{3} \, z R_{edc[a}J_{b]}^{\phantom{b]}cde} \, . \label{biboup}
	\end{align}
\end{subequations}
Assuming that the spacetime geometry is known, so that $(k^a, \! \nabla_a k_b, R_{abcd}, \! \nabla_a R_{bcde})$ are known, and given a physical model for the quadrupole $J^{abcd}$, e.g. Eqs.~\eqref{J_spin} or \eqref{J_tidal}, the relations \eqref{algebraic} are a set of ten \textit{algebraic} equations for the ten unknowns $p^\alpha$ and $S^{\alpha\beta}$. Interestingly, by recalling the Kostant formula \eqref{Kostant} and the expression \eqref{z} for the redshift $z$, the formula \eqref{biboup} appears schematically (getting rid of all tensorial indices and numerical prefactors) as a multipolar identity of the form
\beq
    p \,k + S \,\nabla k + J \,\nabla \nabla k = 0 \, ,
\eeq
while \eqref{bibop} has a similar multipolar structure, with an additional overall covariant derivative. It would be interesting to see if this pattern extends at higher multipolar orders, and to assess whether it carries or not any deeper meaning. Naturally, these equations are closely related to similar formulas established in the context of Dixon's and Harte's formalisms for extended fluid bodies, in presence of an isometry \cite{Di.79,Ha.al.16}.

\section{Conserved quantities}\label{sec:conserved}

In this final section, we explore the various conserved quantities associated with the isometry generated by the helical Killing field \eqref{k}. In particular, given the Lie-dragging along $k^a$ of the 4-velocity, 4-momentum, spin and quadrupole tensor of each particle established in Sec.~\ref{subsec:Lies}, the result \eqref{blibli} implies that any scalar field that is constructed out of the particle's variables $(u^a,p^a,S^{ab},J^{abcd})$ and the spacetime geometry $(g_{ab},k^a,R_{abcd},\dots)$ will be conserved along $\gamma$.

\subsection{Killing energy}\label{subsec:Killing_energy}

For a \textit{generic} Killing vector field $\xi^a$, i.e., for a Killing vector field that does not necessarily satisfy the helical constraint \eqref{k=zu}, the Killing energy of a particle with momentum $p^a$ and spin $S^{ab}$ is defined as
\beq\label{E}
	E_\xi \equiv p^a \xi_a + \frac{1}{2} S^{ab} \nabla_a \xi_b \, .
\eeq
At the dipolar order, this linear combination of $p^a$ and $S^{ab}$ is easily seen to be conserved by substituting the equations of evolution \eqref{EE} with $J^{abcd} = 0$ into the expression for $\dot{E}_\xi$ and by using Killing's equation \eqref{Killing} and the Kostant formula \eqref{Kostant}. Remarkably, the conservation of the Killing energy \eqref{E} holds beyond the dipolar order. Indeed, it can be shown that the scalar \eqref{E} is a constant of motion for an \textit{arbitrary} extended body endowed with an infinite set of multipole moments \cite{EhRu.77,Ha.15,St.15,Di.15}.

In general, however, neither the monopolar contribution, nor the dipolar contribution to the Killing energy \eqref{E}, say
\beq
	E_\xi^{(p)} \equiv p^a \xi_a \quad \text{and} \quad E_\xi^{(S)} \equiv \frac{1}{2} S^{ab}  \nabla_a \xi_b \, ,
\eeq
will be separately conserved. For example, if $t^a$ and $\phi^a$ denote the usual Killing vector fields associated with the stationary and axisymmetry of the Kerr geometry, then a test spinning particle orbiting a spinning black hole has a conserved energy $-E_t$ and a conserved angular momentum $E_\phi$, but the monopolar and dipolar contributions $\{E_t^{(p)},E_t^{(S)}\}$ and $\{E_\phi^{(p)},E_\phi^{(S)}\}$ to $E_t$ and $E_\phi$ are not separately conserved.

However, in our case the helical nature of the Killing field $k^a$ implies the constraint \eqref{k=zu}, from which we readily derive the exact conservation laws
\begin{subequations}
	\begin{align}
		z \dot{E}_k^{(p)} &= \Lik (p^a k_a) = (\Lik p^a) k_a + p^a \Lik k_a = 0 \, , \label{zEmonodot} \\
		2z \dot{E}_k^{(S)} &= \Lik (S^{ab} \nabla_a k_b) = (\Lik S^{ab}) \nabla_a k_b + S^{ab} \Lik \nabla_a k_b = 0 \, , \label{zEdipodot}
	\end{align}
\end{subequations}
where we used Eqs.~\eqref{LieSpin}--\eqref{LieMom} and the identity $\Lik \nabla_a k_b = \nabla_a \Lik k_b = 0$ (see App.~\ref{app:Killing}). So, in our physical setup, the monopolar and dipolar contributions to the Killing energy $E_k$ are \textit{separately} conserved, irrespective of a particular choice of SSC. In particular, by combining Eq.~\eqref{k=zu} with the definition \eqref{defnorms} of the particle's rest mass, the monopolar contribution to the Killing energy is easily seen to coincide with the redshifted rest mass:
\beq\label{pk=mz}
	E_k^{(p)} = p^a k_a = - mz \, .
\eeq
This expression is indeed consistent with the conservation \eqref{zdot} and \eqref{mdot} of $z$ and $m$. \vspace{-0.1cm} The separate conservation of $E_k^{(p)}$ and $E_k^{(S)}$ is a consequence of the constraint \eqref{k=zu} on the helical Killing field, which must be satisfied here because both particles act as a source of spacetime curvature, contrary to the case of a spinning test particle in the Kerr  black hole geometry, for which there exists no relationship (for a generic orbit) between the velocity $u^a$ of the test particle and the Killing fields $t^a|_\gamma$ and $\phi^a|_\gamma$ along the particle's wordline $\gamma$.

\subsection{Other geometrically conserved quantities}\label{subsec:other_conserved}

Equations \eqref{Liu}, \eqref{LieSpin} and \eqref{LieMom} readily imply the Lie-dragging along $k^a$ of the scalar norms \eqref{defnorms}. Combined with the identity \eqref{blibli} we conclude that the rest mass $m = -p^a u_a$, the dynamical mass $\mu^2 = - p^a p_a$ and the spin amplitude $S^2 = \tfrac{1}{2} S^{ab} S_{ab}$ are all conserved along $\gamma$, irrespective of a choice of SSC, i.e.
\beq\label{mdot}
	\dot{m} = 0 \, , \quad \dot{\mu} = 0 \quad \text{and} \quad \dot{S} = 0 \, .
\eeq
Moreover, as shown in App.~\ref{subapp:compendium}, Killing's equation implies the identity $k^c \nabla_c k_a = - \tfrac{1}{2} \nabla_a (k^c k_c)$. When evaluated along $\gamma$, this yields $z^2 \dot{u}_a = \tfrac{1}{2} \nabla_a |k|^2|_\gamma = z \nabla_a |k|$, where we used Eqs.~\eqref{k=zu}--\eqref{zdot}, and the fact that $k^a$ is necessarily timelike in a neighborhood of $\gamma$. The 4-acceleration can thus be expressed in terms of the gradient of the norm of the helical Killing field as
\beq\label{udot}
	\dot{u}_a = \nabla_a \ln{|k|} \, .
\eeq
Contracting with $u^a$ we find that $|k|_\gamma = z$ is conserved along $\gamma$, as established in Sec.~\ref{subsec:Detweiler}.

Moreover, by applying the general result \eqref{comLidot} to the particular case of the 4-velocity $u^a$ of a given particle, while making use of the Lie-dragging \eqref{Liu} of $u^a$, we readily obtain the Lie-dragging along $k^a$ of the 4-acceleration:
\beq\label{Lie-udot}
	\Lik \dot{u}^a = 0 \, .
\eeq
More precisely, the general result \eqref{comLidot} should be applied to an extension $\tilde{u}^a$ of $u^a$ in a neighborhood of $\gamma$, as in \eqref{extensions}. This result can alternatively be derived by taking the Lie derivative along $k^a$ of the expression \eqref{udot} of the 4-acceleration, as $\Lik \nabla_a \ln{|k|} = \nabla_a \Lik \ln{|k|} = 0$. The formula \eqref{Lie-udot} is equivalent to
\beq\label{ddot}
	z \ddot{u}_a = \dot{u}^c \nabla_c k_a \, ,
\eeq
which contracted with the 4-acceleration implies $\dot{u}^a \ddot{u}_a \!=\! 0$, thanks to Killing's equation \eqref{Killing}. Thus, the norm of the acceleration is conserved, in addition to that of the velocity. Moreover, contracting \eqref{ddot} with $u^a$ and using Killing's equation along with the helical constraint \eqref{k=zu} with $z$ constant implies $u^a \ddot{u}_a = - \dot{u}^c \dot{u}_c$, so that
\beq\label{norm-udot}
	\dot{u}^a \dot{u}_a = - \ddot{u}^a u_a = \text{const.}
\eeq
The same argument holds for the rates of change of any Lie-dragged quantity. In particular, $\dot{p}^a \dot{p}_a$, $\dot{S}^a \dot{S}_a$, $\dot{D}^a \dot{D}_a$, $\dot{S}^{ab} \dot{S}_{ab}$ and $\dot{J}^{abcd} \dot{J}_{abcd}$ are all constant along $\gamma$.
 
Finally, using the Kostant formula \eqref{Kostant} together with the constraint \eqref{k=zu}, we can easily show that $\nabla_a k_b$ is conserved along $\gamma$, according to
\beq\label{Dkdot}
	(\nabla_a k_b) \, \dot{} \, \equiv u^c \nabla_c \nabla_a k_b = - u^c R_{abcd} k^d = - z R_{ab(cd)} u^c u^d = 0 \, .
\eeq
As will be shown in \alext{future work}, the conserved norm $|\nabla k|_\gamma$ of the conserved 2-form $\nabla_a k_b|_\gamma$ is very closely related to the precession frequency of the spin vector $S^a$ that was introduced in Sec.~\ref{subsec:SSC}, and which has been used extensively to compare post-Newtonian and gravitational self-force calculations \cite{Do.al.14,BiDa3.14,BiDa2.15,Ak.al.17,Ak.17,Bi.al.18}, and to calibrate effective one-body models \cite{Ka.al.17}.

\acknowledgments

The authors acknowledge the financial support of the Action F\'ed\'eratrice PhyFOG and of the Scientific Council of the Paris Observatory. ALT is grateful to A. Pound for a fruitful suggestion and to R. Porto for an informative email exchange. PR thanks Mathieu Langer for helpful suggestions regarding the structure of this paper and Jan Steinhoff for a fruitful discussion.

\appendix

\section{Summary of conventions and notations}\label{app:conventions}

Our sign conventions are those of \cite{Wal}. In particular, the metric signature is $(-,+,+,+)$, the Riemann tensor satisfies $2 \nabla_{[a} \nabla_{b]} \omega_c = R_{abc}^{\phantom{abc}d} \omega_d$ for any 1-form $\omega_a$, and the Ricci tensor is defined by $R_{ab} = R_{acb}^{\phantom{acb}c}$. Abstract indices are denoted using letters $(a,b,c,\dots)$ from the beginning of the Latin alphabet, while Greek letters $(\alpha,\beta,\gamma,\dots)$ denote tensor components in a given basis. Capital Latin letters $(K,M,N,\dots)$ denote multi-indices of length $k,m,n,\dots$, \alext{as in \cite{BlDa.86},} and the Roman subscript $\ui \in \{1,2\}$ is used to denote the particles in the binary. Throughout this paper we use geometrized units such that $G = c = 1$. For the convenience of the reader, a list of the symbols used most frequently is given in Tab.~\ref{Table}.

\begin{table}[!ht]
    \caption{List of frequently used symbols.}
    \vspace{0.2cm}
	\begin{tabular}{ccc}
		\toprule
		\textbf{Symbol}         & \textbf{Description}          & \textbf{Definition}  \\
		\midrule
		\textbf{Sets}           &                               & \\
		$\mathcal{M}$           & spacetime manifold            & \\ 
		$\scV$                  & a 4-volume in $\mathcal{M}$   & \\ 
		$\gamma$                & particle worldline            & \\
		$x,x'$                  & points in $\mathcal{M}$       & \\
		$y,y'$                  & points on $\gamma$            & \\
		\midrule
		\textbf{Geometry}       &                               & \\ 
		$g_{ab}$                & metric tensor                 & \\ 
        $\nabla_a$              & covariant derivative          & \\ 
		$\varepsilon_{abcd}$    & canonical volume form         & \\
		$R_{abcd}$              & Riemann curvature tensor      & \\ 
		\midrule
		\textbf{Particle}         &                               & \\ 
		$\tau$                  & proper time                   & \\
        $u^a$                   & 4-velocity                    & \\
        $\dot{u}^a$             & 4-acceleration                & \\
		$p^a$                   & 4-momentum                    & \\
		$S^{ab}$                & spin tensor            	    & \\ 
		$J^{abcd}$              & quadrupole tensor             & \\
		$T^{ab}$                & energy-momentum tensor      & \eqref{SET} \\
		$h_{ab}$    & projector orthogonal to $u^a$ & \eqref{h} \\ 
		$\delta_4$              & invariant Dirac distribution  & \eqref{delta4sym2} \\
		\midrule \vspace{0.1cm}
		\textbf{Multipoles}     &                               & \\
		$\mathcal{T}^{ab c_1 \cdots c_\ell}$  & $2^\ell$-pole of $T^{ab}$ (Ansatz)                  & \eqref{AnsatzSETbinary} \\ 
		$\scT^{ab c_1 \cdots c_\ell}$         & $2^\ell$-pole of $T^{ab}$ (normal form)             & \eqref{AnsatzSETnorm} \\ 
		$\tilde{\scT}^{ab c_1 \cdots c_\ell}$ & geodesic extension of $\scT^{ab c_1 \cdots c_\ell}$ & \eqref{extensions} \\
		\midrule
		\textbf{Isometry}       &                               & \\ 
		$\xi^a$                 & generic Killing vector        & \\ 
        $\Lixi$                  & Lie derivative along $\xi^a$    & \eqref{LieOmega} \\
        $k^a$                   & helical Killing vector        & \eqref{k} \\
		$z$                     & redshift parameter            & \eqref{z} \\
        \bottomrule
	\end{tabular}
    \label{Table}
\end{table}

\section{Generalities on bitensors and their Lie derivatives}\label{app:BiDiLie}

In this appendix we shall briefly review the concepts of bitensor, coincidence limit, parallel propagator, invariant Dirac functional and Lie derivative operator. We shall then prove that the Lie derivative of the invariant Dirac distribution along the flow of a Killing field vanishes identically. The reader is referred to, e.g., Ref.~\cite{Syn,Po.al.11} for more details on those notions.

\subsection{Bitensors and coincidence limit}\label{subapp:Bi}

Just like a tensor field is a multilinear map on the points $x$ of a spacetime manifold $\mathcal{M}$, a \textit{bitensor} field is a multilinear map on two points $(x,x') \in \mathcal{M} \times \mathcal{M}$. A generic bitensor will then be denoted as
\beq
    \Omega^{ab\cdots}_{\phantom{ab\cdots}a'b'\cdots}(x,x') \, ,
\eeq
where the abstract indices $a,b,c,\dots$ and $a',b',c',\dots$ refer to the points $x$ and $x'$, respectively. Two examples of bitensors are used in this paper: the parallel propagator $g^{a}_{\phantom{a}a'}(x,x')$ and the invariant Dirac functional $\delta_4(x,x')$, both defined below.

An important operation for bitensors is the \textit{coincidence limit}, which consists in evaluating a bitensor at the same point. It is defined as 
\beq\label{coincidence}
    \bigl[ \Omega^{ab\cdots}_{\phantom{ab\cdots}a'b'\cdots} \bigr](x) \equiv \lim_{x'\rightarrow x} \Omega^{ab\cdots}_{\phantom{ab\cdots}a'b'\cdots}(x,x')\, .
\eeq
The coincidence limit of a bitensor is thus an ordinary tensor field. We will assume that this coincidence limit always exist and is independent of the direction in which $x'$ approaches $x$. For more details regarding bitensors and the coincidence limit, see e.g. Ref.~\cite{Po.al.11}.

\subsection{Parallel propagator}\label{subapp:paral}

An important example of bitensor is the parallel propagator. If the points $x$ and $x'$ are ``close enough,'' i.e., if $x'$ is in a normal neighborhood of $x$, then there is a unique geodesic segment $\lambda$ that joins them. On this geodesic segment, we introduce an orthonormal tetrad $(e^a_A)$ that is parallel-transported on $\lambda$, where the subscript $A \in \{0,1,2,3\}$ labels the vectors of the basis. By definition, this tetrad obeys the orthonormality and completion relations
\beq \label{tetrad}
    g_{ab} e^a_A e^b_B = \eta_{AB} \quad \text{and} \quad g_{ab} = \eta_{AB} e^A_a e^B_b \, ,
\eeq
where the Minkowski metric $\eta_{AB} = \text{diag}\,(-1,1,1,1)$  is used to lower the Greek indices, and its inverse $\eta^{AB}$ to raise them. The 1-form $e_a^A$ is defined by metric duality as $e_a^A \equiv \eta^{AB} g_{ab} e^b_B$.

Next, we introduce a generic vector field $v^a$ defined on $\lambda$. At any point $z\in\lambda$, this vector field can be expanded with respect to the tetrad $(e^a_A)$, according to
\beq \label{exp}
    v^a(z) = v^A(z) e^a_A(z) \, , \quad \text{where} \quad v^A \equiv v^a e^A_a \, .
\eeq
Now we make the following remark: if $v^a$ is parallely transported along $\lambda$, then it is clear from \eqref{exp} that the tetrad components $v^A$ remain constant along $\lambda$, and thus have the same numerical value at $z=x$ and at $z=x'$. By substituting the definition given in the right-hand side of Eq.~\eqref{exp} in each side of the equality $v^A(x)=v^A(x')$, and by using the orthogonal properties of the tetrad \eqref{tetrad}, we obtain
\beq \label{defpara}
    v^a(x) = g^a_{\phantom{a}a'}(x,x') v^{a'}(x') \, , \quad \text{where} \quad g^a_{\phantom{a}a'}(x,x')\equiv e_A^a(x) e^A_{a'}(x') \, .
\eeq
The bitensor $g^a_{\phantom{a}a'}(x,x')$ is the so-called \textit{parallel propagator} from $x'$ to $x$. It takes the vector $v^{a}$ at the point $x'$ and extends it by parallel transport to the point $x$. As long as the underlying geodesic is unique, this extension is unique as well. The formula \eqref{defpara} can be generalized to a generic tensor field of arbitrary rank. The parallel propagator \eqref{defpara} is used in Sec.~\ref{subsec:Lieconstraints} to extend the multipoles off the worldline of each quadrupolar particle. 

\subsection{Invariant Dirac distribution}

The gravitational skeleton model reviewed in Sec.~\ref{sec:skeleton} and used in Secs.~\ref{sec:colinear}--\ref{sec:conserved} relies crucially on a 4-dimensional, covariant generalization of the ordinary, noncovariant Dirac distribution. In four spacetime dimensions, the \textit{invariant} Dirac functional $\delta_4(x,x')$ is the distributional biscalar defined by the relations \cite{Po.al.11}
\beq\label{delta4}
	\int_{\scV} f(x) \, \delta_4 (x,x') \, \ud V = f(x') \quad \text{and} \quad \int_{\scV'} f(x') \, \delta_4 (x,x') \, \ud V' = f(x) \, ,
\eeq
where $f$ is a smooth scalar field (a test function), $\scV$ and $\scV'$ any four-dimensional regions of spacetime that contain the points $x'$ and $x$, respectively, and $\ud V = \sqrt{-g} \, \ud^4 x$ is the invariant volume element, with $g$ the determinant of the metric tensor $g_{ab}$ in a given coordinate basis. The definition \eqref{delta4} ensures that $\delta_4$ is symmetric in its arguments,
\beq\label{delta4sym}
	\delta_4(x,x') = \delta_4(x',x) \, ,
\eeq
such that it depends necessarily on the \textit{difference} of the events' coordinates. More precisely, given a coordinate system $(x^\alpha)$, one can easily show that \cite{Po.al.11}
\beq\label{delta4sym2}
	\delta_4(x,x') = \prod_{\alpha=0}^3 \frac{\delta(x^\alpha-x'^\alpha)}{\sqrt{-g}} \, ,
\eeq
where $\delta$ is the ordinary, noncovariant Dirac distribution, such that $\int_{\mathbb{R}} \phi(t) \delta(t) \, \ud t = \phi(0)$ for any test function $\phi$. Together with the consequence $\nabla_{\!a} \, g = 0$ of metric compatibility, the explicit formula \eqref{delta4sym2} implies the property
\beq\label{grad_delta4}
    (\nabla_a + \nabla_{a'}) \, \delta_4(x,x') = 0 \, .
\eeq
Finally, by recalling the notation \eqref{coincidence} for the coincidence limit where $x' \to x$, an important distributional identity satisfied by $\delta_4$, valid for any bitensor $\Omega^{ab\cdots}_{\phantom{ab\cdots}a'b'\cdots}$, is
\beq\label{azerty}
	\Omega^{ab\cdots}_{\phantom{ab\cdots}a'b'\cdots}(x,x') \, \delta_4(x,x') = \bigl[ \Omega^{ab\cdots}_{\phantom{ab\cdots}a'b'\cdots} \bigr] \, \delta_4(x,x') \, .
\eeq

\subsection{Lie derivative operator}

Our derivation of the relations \eqref{k=zu}, \eqref{Liu}, \eqref{LieQuad}, \eqref{LieSpin} and \eqref{LieMom} relies on the invariance \eqref{LieT} of the quadrupolar energy-momentum tensors \eqref{SET} and \eqref{AnsatzSETnorm} along the integral curves of the helical Killing vector \eqref{k}. Since these tensor fields involve the (distributional) bitensor $\delta_4(x,x')$, we require a generalization to bitensor fields of the ordinary definition of the Lie derivative of a smooth tensor field.

Let $\phi_\epsilon$ denote a one-parameter group of diffeomorphism generated by a vector field $\xi^a(x)$. The ``push-forward'' $\phi^*_\epsilon$ can then be used to carry any smooth bitensor field $\Omega^{ab\cdots}_{\phantom{ab\cdots}a'b'\cdots}(x,x')$ along the flow of $\xi^a$, by acting independently on both spacetime points $x$ and $x'$. By analogy with the definition of the Lie derivative of a smooth tensor field, we define the Lie derivative $\Lixi$ along $\xi^a$ of a smooth bitensor field as \cite{Wal}
\beq\label{LieOmega}
	\Lixi \Omega^{ab\cdots}_{\phantom{ab\cdots}a'b'\cdots} \equiv \lim_{\epsilon \to 0} \, \frac{1}{\epsilon} \left[ \phi^*_{-\epsilon}\Omega^{ab\cdots}_{\phantom{ab\cdots}a'b'\cdots} - \Omega^{ab\cdots}_{\phantom{ab\cdots}a'b'\cdots} \right] ,
\eeq
where all bitensor appearing in \eqref{LieOmega} are evaluated at the same combination $(x,x')$ of points. For a generic biscalar $S(x,x')$, this general definition reduces to
\beq\label{LieS}
	\Lixi S(x,x') = \xi^a \nabla_a S(x,x') + \xi^{a'} \nabla_{a'} S(x,x') \, .
\eeq
This definition coincides with that used by Harte \cite{Ha.12}, who defines the Lie derivative of any bitensor as acting independently and linearly on each spacetime point.

\subsection{Lie derivative of the invariant Dirac functional}

Finally, we wish to establish a formula for the Lie derivative $\Lixi \delta_4$ along a vector field $\xi^a$ of the invariant Dirac distribution $\delta_4$. By applying the definition \eqref{LieS} of the Lie derivative to the distributional biscalar \eqref{delta4sym2}, and by using the property \eqref{grad_delta4}, the Lie derivative of the invariant Dirac distribution along a smooth vector field $\xi^a$ reads
\beq\label{ploup}
	\Lixi \delta_4(x,x') = \bigl( \xi^a - \xi^{a'} \bigr) \nabla_a \delta_4(x,x') \, .
\eeq
This form can be further simplified by integrating $\Lixi \delta_4$ against an arbitrary ``test function.'' Indeed, for any smooth scalar field $f$ with compact support, the formula \eqref{ploup} implies
\beq\label{tutu}
	\int_\scV f(x) \, \Lixi \delta_4 (x,x') \, \ud V = \int_{\scV} \nabla_a \bigl( f \bigl( \xi^a - \xi^{a'} \bigr)\delta_4 \bigr) \, \ud V - \int_{\scV} \Bigl( \nabla_a f \bigl( \xi^a - \xi^{a'} \bigr) + f \nabla_a \xi^a \Bigr) \, \delta_4 \, \ud V \, ,
\eeq
where we integrated by parts and used $\nabla_a \xi^{a'} = 0$. The first integral in the right-hand side can be converted into a surface integral by applying Stokes' theorem, and easily shown to vanish thanks to the distributional identity \eqref{azerty} and the coincidence limit $\bigl[ \xi^a - \xi^{a'} \bigr] = 0$:
\beq\label{tutu2}
	\int_{\scV} \nabla_a \bigl( f \bigl( \xi^a - \xi^{a'} \bigr)\delta_4 \bigr) \, \ud V = \oint_{\partial\scV} \bigl[ f \bigl( \xi^a - \xi^{a'} \bigr) \bigr] \delta_4 \, \ud \Sigma_a = 0 \, .
\eeq
Moreover, by using the distributional identity \eqref{azerty}, the coincidence limit $\bigl[ \xi^a - \xi^{a'} \bigr] = 0$ and the defining property \eqref{delta4} of the invariant distribution $\delta_4$, the second term in the right-hand side of Eq.~\eqref{tutu} simply reads
\beq\label{tutu3}
	\int_{\scV} \Bigl( \bigl[ \nabla_a f \bigl( \xi^a - \xi^{a'} \bigr) \bigr] + f \nabla_a \xi^a \Bigr)(x) \, \delta_4(x,x') \, \ud V = \bigl( f \nabla_a \xi^a \bigr) (x') \, .
\eeq
Hence, by substiting \eqref{tutu2} and \eqref{tutu3} into \eqref{tutu}, while using the formula $\nabla_a \xi^a = \frac{1}{2} g^{ab} \Lixi g_{ab}$, we obtain the distributional identity
\beq\label{Lie-delta4}
	\Lixi \delta_4(x,x') = - \frac{1}{2} \, \delta_4(x,x') \, g^{ab}(x) \Lixi g_{ab}(x) \, .
\eeq
This agrees with Eq.~(136) in \cite{Ha.15}, where the same definition of the Lie derivative acting on bitensors was introduced.

In the case where $\xi^a$ is a Killing vector field (see App.~\ref{app:Killing} below), Eq.~\eqref{Lie-delta4} shows that the Dirac functional $\delta_4(x,x')$ is invariant along the integral curves of a Killing field. In particular, for the helical Killing field \eqref{k} considered in this work, the distributional identity \eqref{Lie-delta4} implies that
\beq\label{Liek-delta4}
	\Lik \delta_4(x,y) = 0
\eeq
for any point $y \in \gamma$. This result was used in Sec.~\ref{sec:colinear} to establish the helical constraint \eqref{k=zu}, and in Sec.~\ref{sec:Lips} to derive the Lie-dragging along $k^a$ of the velocity $u^a$, momentum $p^a$, spin $S^{ab}$ and quadrupole $J^{abcd}$ of each quadrupolar particle, Eqs.~\eqref{Liu}, \eqref{LieQuad}, \eqref{LieSpin} and \eqref{LieMom}.

\section{Theorems} \label{app:thm}

In this appendix, we shall review Tulczyjew's two theorems \cite{Tu.59}, which play a central in any work that relies on the gravitational skeleton formalism. The first theorem ensures the existence and unicity of the normal form of a tensor expressed as a distributional multipolar expansion. The second theorem gives a necessary and sufficient condition for such a distributional multipolar expansion to vanish: that the multipoles of its normal form all vanish identically. Finally, we give a proof of the proposition \eqref{claim}, which was used to derive the key helical constraint \eqref{k=zu}. \alext{In what follows, we will use the notation $[\![p,q]\!]$ to denote the set of  integers between any two given integers $(p,q) \in \mathbb{N} \times \mathbb{N}$ with $p < q$.}

\subsection{Tulczyjew's first theorem} \label{app:thm1}

First we introduce some notation. Let $Y^M \equiv Y^{a_1 \cdots a_m}$ denote a contravariant tensor field of rank $m\in\mathbb{N}$. We assume that its support is restricted to a worldline $\gamma$ with proper time $\tau$ and unit tangent $u^a$, and that it can be written as a distributional multipolar expansion of order $n\in\mathbb{N}$. Therefore, at any point $x\in\mathcal{M}$ we have
\beq\label{tulcz_0}
	Y^M(x) = \sum_{k=0}^n \nabla_K \int_{\gamma} \mathcal{Y}^{MK}(y) \, \delta_4(x,y) \, \ud \tau \, ,
\eeq
where $y\in\gamma$ and $(\mathcal{Y}^{MK})_{k\in[\![0,n]\!]}$ is a collection of $n+1$ multipoles, i.e., contravariant tensors of rank $m+k$ defined along $\gamma$. We introduced the notations $\nabla_K \equiv \nabla_{c_1 \cdots c_k}$ and $\mathcal{Y}^{MK} \equiv \mathcal{Y}^{M c_1 \cdots c_k}$ for $k \geqslant 1$, while $\nabla_K=\text{id}$ and $\mathcal{Y}^{MK}=\mathcal{Y}^{M}$ for $k=0$. We may now state the first theorem.
\begin{theorem}\label{thm1}
For any given $(m,n)\in\mathbb{N}\times\mathbb{N}$, let $Y^M$ be defined as in Eq.~\eqref{tulcz_0}. Then there exists a collection of multipoles $(\scY^{MK})_{k\in[\![0,n]\!]}$ that are (i) symmetric with respect to any pair of indices of the multi-index $K$, (ii) orthogonal to $u^a$ with respect to any index of $K$, and (iii) such that
\beq\label{expnorm}
    Y^M(x) = \sum_{k=0}^n \nabla_K \int_\gamma \scY^{MK}(y) \, \delta_4(x,y) \, \ud \tau \, .
\eeq
Moreover, the multipolar expansion \eqref{expnorm} is unique and the multipoles $(\scY^{MK})_{k\in[\![0,n]\!]}$ can be written explicitly in terms of the multipoles $(\mathcal{Y}^{MK})_{k\in[\![0,n]\!]}$ of \eqref{tulcz_0}. Equation \eqref{expnorm} is referred to as the normal form of $Y^M$.
\end{theorem} 
Unicity of the normal form is straightforward once we have Thm.~\ref{thm2} below. For the existence, we construct in App.~\ref{app:reduc} below the explicit normal form associated with the quadrupolar ($n=2$) gravitational skeleton of the generic tensor $Y^M$. In particular, the multipoles $\scY^{MK}$ of the quadrupolar normal form are given in terms of the multipoles $\mathcal{Y}^{MK}$ in Eqs.~\eqref{scoubidoubidou}. For the existence of the normal form when $n > 2$, see e.g. Ref.~\cite{StPu.10} and references therein.

\subsection{Extension of Tulczyjew's second theorem} \label{app:thm2}

Again, we first introduce some notation. Let $p\in\mathbb{N}^\star$ and let $(Y^M_\ui)_{\ui\in[\![1,p]\!]}$ denote a collection of $p$ contravariant tensor fields of rank $m\in\mathbb{N}$. We assume that the support of each $Y^M_\ui$ is restricted to a worldline $\gamma_\ui$ with proper time $\tau_\ui$ and unit tangent $u^a_\ui$, and that it can be expressed as a distributional multipolar expansion of order $n\in\mathbb{N}$. Therefore, at any point $x\in\mathcal{M}$, we have
\beq\label{tulcz_1}
	Y^M_\ui(x) = \sum_{k=0}^n \nabla_K \int_{\gamma_\ui} \scY^{MK}_\ui(y_\ui) \, \delta_4(x,y_\ui) \, \ud \tau_\ui \, ,
\eeq
where, for each $\ui \in [\![1,p]\!]$, $(\scY^{MK}_\ui)_{k\in[\![0,n]\!]}$ is a collection of $n+1$ multipoles, i.e., contravariant tensors of rank $m+k$ defined along $\gamma$. We may now state the second theorem.
\begin{theorem}\label{thm2}
For any given $(m,n,p)\in\mathbb{N}\times\mathbb{N}\times\mathbb{N}^*$, let $(Y^M_\ui)_{\ui \in[\![1,p]}$ be a collection of $p$ tensors defined as in Eq.~\eqref{tulcz_1}, and let $Y^M \equiv \sum_\ui Y^M_\ui$ denote their sum. The following result holds. If for all $(\ui,k)\in [\![1,p]\!]\times[\![0,n]\!]$, $\scY^{MK}_\ui$ is symmetric with respect to any pair of indices of the multi-index $K$ and is orthogonal to $u^a_\ui$ with respect to any index of $K$, then
\beq
    Y^M = 0 \quad \Longleftrightarrow \quad \forall \, (\ui,k) \in [\![1,p]\!] \times [\![0,n]\!] \, , \, \, \scY^{MK}_\ui = 0 \, .
\eeq
\end{theorem}
Clearly, if all the multipoles $\scY^{MK}_\ui$ vanish, then $Y^M_\ui=0$ by \eqref{tulcz_1}, and the sum $Y^M=\sum_\ui Y^M_\ui$ vanishes as well. The heart of the proof therefore resides in showing that if $Y^M_\ui$ is in normal form, then $Y^M=0$ implies $\scY^{MK}_\ui=0$ for all $(\ui,k) \in [\![1,p]\!]\times[\![0,n]\!]$. See Refs.~\cite{Tu.59,Tr.02,StPu.10} and references therein for more details. 

Importantly, in the literature this result is proven for a single multipolar particle, whereas Thm.~\ref{thm2} is stated for an arbitrary number $p \in \mathbb{N}^*$ of particles. However, we now argue that the multi-particle case can easily be reduced to the single-particle case. Indeed, the general idea behind the proof for a single particle is the following: take an arbitrary rank-$m$ tensor $Z_M$, whose compact support $\scV$ intersects the worldline $\gamma$ of the particle. Contract $Z_M$ with $Y^M$, given as a multipolar expansion in normal form \eqref{expnorm}, and integrate over $\scV$. The goal is then to show that this integral vanishes for every $Z_M$ only if $\scY^{MK}=0$ for all $k \in [\![0,n]\!]$. Since this holds for any compact support $\scV$, the vanishing of $\scY^{MK}$ must hold for any portion of $\gamma$, and thus on all of $\gamma$. Now, if there are $p \in \mathbb{N}^*$ multipolar particles, one may choose the volume $\scV$ such that it intersects \textit{only one} of the $p$ worldlines, say $\gamma$, and proceed with the single-particle proof, as summarized above.

\subsection{Proof of proposition \texorpdfstring{\eqref{claim}}{}} \label{subapp:detailsfab}

We now give a proof of the proposition \eqref{claim}, which was used in Sec.~\ref{subsec:quadKilling} to derive the colinearity \eqref{k=zu} of $k^a$ and $u^a_\ui$ along the worldline $\gamma_\ui$ of the $\ui$-th particle. For clarity's sake we will drop the subscript $\ui$, as the proof holds for any of the two particles of the binary system. The proposition \eqref{claim} is an implication which is most easily proven by contraposition.

\begin{figure}[t!]
    \begin{center}
    	\includegraphics[width=0.4\linewidth]{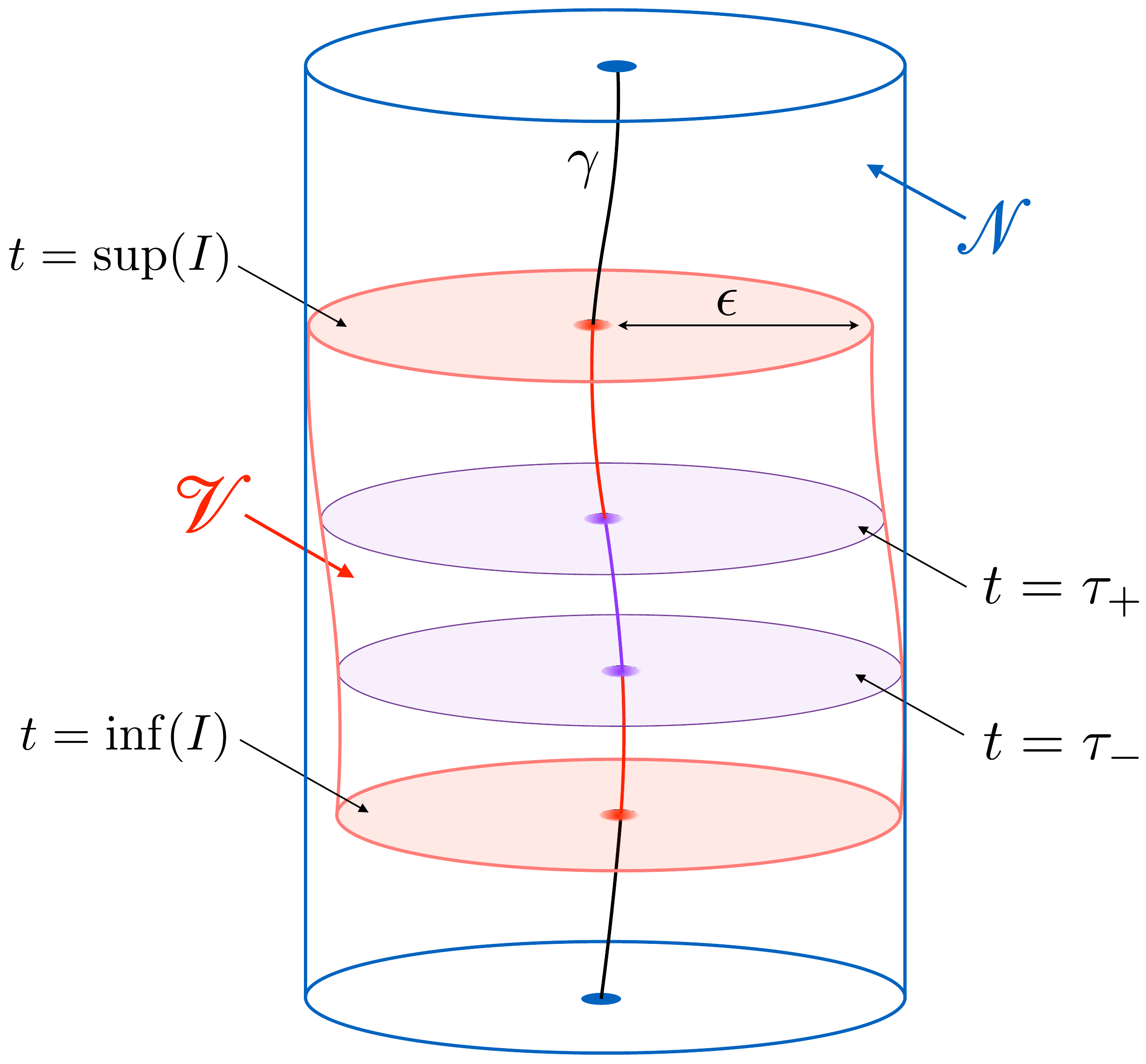}
        \caption{The geometrical setup used in App.~\ref{subapp:detailsfab} to prove the proposition \eqref{claim}.}
        \label{fig:setup}
    \end{center}
\end{figure} 

First, we introduce Fermi coordinates $(t,x^i)$ in a neighborhood $\scN \!\subset\! \mathcal{M}$ of the worldline $\gamma$. Using Fermi coordinates, $\gamma$ is parameterized by the proper time $\tau \in \RR$ according to $(t,x^i)=(\tau,0,0,0)$. Let $I\subset\RR$ be a finite interval and $\gamma_I$ be the finite portion of $\gamma$ parameterized by $\tau \in I$. We also set $\epsilon>0$ and let $\scV$ denote the 3-cylinder of Fermi coordinate radius $\epsilon$ that surrounds $\gamma_I$. We assume that $\epsilon$ is small enough such that $\scV \subseteq \scN$. This geometric setup is depicted on Fig.~\ref{fig:setup}. Finally, for given tensor fields $(\tilde{\scT}^{ab}, \tilde{\scT}^{abc},\tilde{\scT}^{abcd})$ defined in Eq.~\eqref{extensions}, we let $\FF \equiv \{ \tilde{\scT}^{ab}f_{ab} + \tilde{\scT}^{abc}\nabla_cf_{ab} + \tilde{\scT}^{abcd}\nabla_{cd}f_{ab} \,,f_{ab}\in C^{\infty}_\scV\}$, with $C^{\infty}_\scV$ the set of tensor fields with compact support $\scV$ that are smooth on the interior $\scV^\circ$. 

\subsubsection{\texorpdfstring{Proof that $z=\text{cst}$ along $\gamma$}{}}

Since all the scalar fields are evaluated along $\gamma$ in the following integrals, we view them as functions of the proper time $\tau$. We first prove by contraposition the part of the proposition \eqref{claim} that implies that $\dot{z}=0$ along $\gamma$, i.e., we show that the following implication is true:
\beq\label{claim1}
    \exists \, \tau_0\in I \, , \, \, \dot{z}(\tau_0)\neq 0 \quad \Longrightarrow \quad \exists \, f \in \FF \, , \, \,  \int_\RR \bigl( z \dot{f} + w^a\nabla_a f \bigr) \, \ud \tau \neq 0 \, .
\eeq
Let $\tau_0\in I$ be as in Eq.~\eqref{claim1}. Since $\tau\mapsto\dot{z}(\tau)$ is continuous, there exists a neighborhood of $\tau_0$, say $]\tau_-,\tau_+[ \; \subset I$, such that $\forall \tau \in \; ]\tau_-,\tau_+[\, ,\dot{z}(\tau)\neq 0$ and is of constant sign. Now consider the scalar field $f$ defined on $\scV^\circ$ by
\beq\label{deff}
    f(t,x^i) \equiv
    \begin{cases}
	    \; \exp{ \bigl( [(t-\tau_-)(t-\tau_+)]^{-1} \bigr)} \quad &\text{if} \;\; t \in \; ]\tau_-,\tau_+[ \; , \\
	    \; 0 \quad &\text{if} \;\; t \notin \; ]\tau_-,\tau_+[ \; ,
	\end{cases}
\eeq
and $f\equiv 0$ elsewhere. We claim that such $f$ verifies the right-hand side of \eqref{claim1}.

Indeed, in the Fermi coordinate system, the 4-velocity $u^a$ has components $u^\alpha = (1,0,0,0)$, so that $w^au_a=0$ implies that $w^a$ has components $w^\alpha = (0,w^i)$. Hence $w^a\nabla_a f=w^i\partial_i f =0$ since $f$ does not depend on $x^i$. Moreover, the function $\tau\mapsto f(\tau)$ is smooth on $\mathbb{R}$ and vanishes for $\tau \notin \; ]\tau_-,\tau_+[$. Consequently $\int_\gamma (z \dot{f} +w^a\nabla_af) \, \ud \tau = \int^{\tau_+}_{\tau_-} z(\tau) \dot{f}(\tau) \, \ud \tau$, and an integration by parts gives
\beq\label{intfirst}
    \int_\gamma \bigl( z \dot{f} + w^a \nabla_a f \bigr) \, \ud \tau = - \int^{\tau_+}_{\tau_-} \dot{z}(\tau) f(\tau) \, \ud \tau \, .
\eeq
But the integral on the right-hand side of \eqref{intfirst} cannot vanish, as $\dot{z}$ is nonzero with constant sign over $]\tau_-,\tau_+[$ by assumption, and $f(\tau) > 0$ for all $\tau \in \; ]\tau_-,\tau_+[$. Therefore, $f$ as defined in Eq.~\eqref{deff} verifies the proposition \eqref{claim1}, provided that it belongs to the set $\FF$.

To establish that $f \in \FF$, consider the tensor $f_{ab}\equiv\phi g_{ab}$, where the scalar field $\phi$ is defined over $\scV^\circ$ by
\beq\label{defphi}
    \phi(t,x^i) \equiv
    \begin{cases}
		\; \exp{\bigl( [(t-\tau_-)(t-\tau_+)]^{-1} \bigr)} \, \Phi(t)^{-1} \quad &\text{if} \;\; t \in \; ]\tau_-,\tau_+[ \; , \\
		\; 0 \quad &\text{if} \;\; t \notin \; ]\tau_-,\tau_+[ \; ,
	\end{cases}
\eeq
and $\phi\equiv 0$ elsewhere, where $\Phi(t) \equiv g_{ab} \tilde{\scT}^{ab}$, the latter being evaluated at the point $(t,0,0,0)$. With $f_{ab}=\phi g_{ab}$ and $\phi$ given in \eqref{defphi}, one can readily check that $f= \tilde{\scT}^{ab}f_{ab}-\tilde{\scT}^{abc}\nabla_cf_{ab}+\tilde{\scT}^{abcd}\nabla_{cd}f_{ab}$, for $f$ given in Eq.~\eqref{deff}. The computation involves (i) the metric compatibility $\nabla_c g_{ab}=0$, (ii) the independence of $\phi$ with respect to $x^i$, (iii) the fact that the Christoffel symbols $\Gamma^t_{\phantom{t}ij}|_\gamma$ vanish in Fermi coordinates, and (iv) the normal form of the tensors $\tilde{\scT}^{ab}$, $\tilde{\scT}^{abc}$ and $\tilde{\scT}^{abcd}$.

\subsubsection{\texorpdfstring{Proof that $w^a=0$ along $\gamma$}{}}

Having proven that $z$ is constant along $\gamma$, it is clear that for any compactly supported $f$, we have $\int_\gamma z\dot{f} \, \ud \tau = 0$. Consequently, we will now establish that $w^a=0$ along $\gamma$ by proving the following proposition:
\beq\label{claim2}
    \exists \, \tau_0\in I \, , \, \, w^a(\tau_0) \neq 0 \quad \Longrightarrow \quad \exists \, f \in \FF \, , \, \,  \int_\gamma w^a\nabla_a f \, \ud \tau \neq 0 \, .
\eeq
As noted before, in the Fermi coordinate system we have $w^\alpha \!=\! (0,w^i)$ so that $w^a\nabla_a f \!=\! w^i \partial_i f$. Because $w^i$ is continuous and $w^i(\tau_0)\neq 0$, there exists a neighborhood $]\tau_-,\tau_+[ \; \subset I$ of $\tau_0$ such that \textit{at least} one component of $w^i$, say $w^1$, is nonzero and of constant sign over $]\tau_-,\tau_+[$. Now consider the following scalar field defined on $\scV^\circ$:
\beq\label{deff_bis}
    f(t,x^i) \equiv
    \begin{cases}
	    \; x^1 \exp{\bigl( [(t-\tau_-)(t-\tau_+)]^{-1} \bigr)} \quad &\text{if} \;\; t \in \; ]\tau_-,\tau_+[ \; , \\
	    \; 0 \quad &\text{if} \;\; t \notin \; ]\tau_-,\tau_+[ \; ,
	\end{cases}
\eeq
and $f\equiv0$ elsewhere. Because $f$ does not depend on $x^2$ and $x^3$, the integral in the right-hand side of Eq.~\eqref{claim2} is simply 
\beq
	\int_\gamma w^a \nabla_a f \, \ud \tau = \int_{\tau_-}^{\tau_+} w^1(\tau) \, \exp{\bigl( [(\tau-\tau_-)(\tau-\tau_+)]^{-1} \bigr)} \, \ud \tau \, .
\eeq
As earlier this integral does not vanish since, by assumption, $w^1(\tau)$ is nonzero and of constant sign over $]\tau_-,\tau_+[$. Therefore, the scalar field \eqref{deff_bis} verifies Eq.~\eqref{claim2} provided that it belongs to $\FF$. Once again, let us consider the tensor field $f_{ab} \equiv \phi g_{ab}$, with the scalar field $\phi$ now defined on $\scV^\circ$ by 
\beq\label{defphi_bis}
    \phi(t,x^i) \equiv
    \begin{cases}
		\; \frac{1}{6}(x^1)^3 \, \exp{\bigl( [(t-\tau_-)(t-\tau_+)]^{-1} \bigr)} \, \Phi(t)^{-1} \quad &\text{if} \;\; t \in \; ]\tau_-,\tau_+[ \; , \\
		\; 0 \quad &\text{if} \;\; t \notin \; ]\tau_-,\tau_+[ \; ,
	\end{cases}
\eeq
and $\phi\equiv 0$ elsewhere, where this time $\Phi(t)\equiv g_{\alpha\beta} \bigl( \tfrac{1}{6} (x^1)^2\tilde{\scT}^{\alpha\beta}-\tfrac{1}{2}x^1\tilde{\scT}^{\alpha\beta1}+\tilde{\scT}^{\alpha\beta11} \bigr)$, the latter being evaluated at the point $(t,0,0,0)$. With $f_{ab}=\phi g_{ab}$ and $\phi$ given in Eq.~\eqref{defphi_bis}, one can readily check that $f= \tilde{\scT}^{ab}f_{ab}- \tilde{\scT}^{abc}\nabla_cf_{ab}+\tilde{\scT}^{abcd}\nabla_{cd}f_{ab}$, for $f$ given in \eqref{deff_bis}. This time, the computation involves (i) the metric compatibility $\nabla_c g_{ab}=0$, (ii) the independence of $\phi$ with respect to the coordinates $x^2$ and $x^3$, (iii) the fact that the Christoffel symbols $\Gamma^\alpha_{\phantom{a}ij}|_\gamma$ vanish in Fermi coordinates, and (iv) the normal form of the tensors $\tilde{\scT}^{ab}$, $\tilde{\scT}^{abc}$ and $\tilde{\scT}^{abcd}$.

\section{Normal form of a quadrupolar gravitational skeleton} \label{app:reduc}

In this appendix, we shall detail the computations that lead to the unique normal form associated with the quadrupolar gravitational skeleton of a generic tensor field. This normal form can for instance be used to derive the equations of evolution for the momentum and spin of a \textit{dipolar} particle, i.e. Eq.~\eqref{EE} with $J^{abcd}=0$, or to obtain the Lie-dragging constraints \eqref{constraints} for a quadrupolar particle.

\subsection{A useful formula}

Before deriving this normal form, we first prove a simple formula that will turn out crucial in order to carry out the following computations. Let $\bm{T}$ denote a generic tensor field defined along the worldline $\gamma$ with unit tangent $u^a$. Then we have
\begin{align}
    \nabla_a \int_\gamma \bm{T}(y') \, u^{a'}(y') \, \delta_4(x,y') \, \ud \tau
    &= \int_\gamma \bm{T}(y') \, u^{a'}(y') \nabla_a \delta_4(x,y') \, \ud \tau \nonumber \\
    &= - \int_\gamma \bm{T}(y') \, u^{a'}(y') \nabla_{a'} \delta_4(x,y') \, \ud \tau \nonumber \\
    &= - \int_\gamma \bigl[ \bm{T}(y') \, \delta_4(x,y') \bigr] \, \dot{} \; \ud \tau + \int_\gamma \dot{\bm{T}}(y') \, \delta_4(x,y') \, \ud \tau \, ,
\end{align}
where we used the fact that the covariant derivative $\nabla_a$ acts on points $x\in\mathcal{M}$ but not on points $y'\in\gamma$ in the first equality, the property \eqref{grad_delta4} of the invariant Dirac functional in the second equality, and we integrated by parts in the third and last equality. Assuming that $\bm{T}$ vanishes as $\tau \to \pm \infty$ to discard the boundary terms, we conclude that for any tensor field $\bm{T}$ defined along $\gamma$,
\beq\label{magic}
    \nabla_a \int_\gamma \bm{T} \, u^a \, \delta_4 \, \ud \tau = \int_\gamma \dot{\bm{T}} \, \delta_4 \, \ud \tau \, .
\eeq

\subsection{Derivation of the normal form}

We now turn to the derivation of the normal form at quadrupolar order. Let $Y^M$ denote a generic tensor field of rank $m\in\mathbb{N}$, expressed as a gravitational skeleton at quadrupolar order, i.e., Eq.~\eqref{tulcz_0} with $n=2$, such that 
\beq\label{start}
    Y^M = \int_\gamma \mathcal{Y}^{M} \delta_4 \, \ud \tau 
+ \nabla_a \int_\gamma \mathcal{Y}^{Ma} \delta_4 \, \ud \tau 
+ \nabla_{ab} \int_\gamma \mathcal{Y}^{Mab} \delta_4 \, \ud \tau \, ,
\eeq
with $\mathcal{Y}^{M}$, $\mathcal{Y}^{Ma}$ and $\mathcal{Y}^{Mab}$ the monopole, dipole and quadrupole of $Y^M$, respectively. From Thm.~\ref{thm1} the first term in \eqref{start} is already in normal form. For the second term, we perform an orthogonal decomposition of $\mathcal{Y}^{Ma}$ with respect to the index $a$ by means of the projector \eqref{h} orthogonal to the 4-velocity $u^a$, namely $\mathcal{Y}^{Ma} = \mathcal{Y}^{M\hat{a}} - \mathcal{Y}^{Mu}u^a$. (Recall the notations introduced below Eqs.~\eqref{normalformquad}.) Using the formula \eqref{magic} then gives 
\beq\label{Deu}
    \nabla_a \int_\gamma \mathcal{Y}^{Ma} \delta_4 \, \ud \tau = \nabla_a \int_\gamma \mathcal{Y}^{M\hat{a}} \delta_4 \, \ud \tau - \int_\gamma \bigl(\mathcal{Y}^{Mu}\bigr)\,\dot{}\;\delta_4 \, \ud \tau \, .
\eeq
Regarding the third term on the right-hand side of Eq.~\eqref{start}, we start again by performing an orthogonal decomposition of the integrand, yielding $\mathcal{Y}^{Mab}=\mathcal{Y}^{M\hat{a}\hat{b}}-\mathcal{Y}^{Mu\hat{b}}u^a-\mathcal{Y}^{Mau}u^b$. Substituting this decomposition into the integral and using the formula \eqref{magic}, we obtain
\beq \label{troy}
    \nabla_{ab} \int_\gamma \mathcal{Y}^{Mab} \delta_4 \, \ud \tau = \nabla_{ab} \int_\gamma \mathcal{Y}^{M\hat{a}\hat{b}} \delta_4 \, \ud \tau
- \nabla_{ab} \int_\gamma \mathcal{Y}^{Mu\hat{b}} u^a \delta_4 \, \ud \tau 
- \nabla_a \int_\gamma \bigl(\mathcal{Y}^{Mau}\bigr)\,\dot{}\;\delta_4 \, \ud \tau \, .
\eeq
We shall now consider those three terms successively. 

We begin with the first term of \eqref{troy}. We split the second covariant derivative into its symmetric and antisymmetric part, $\nabla_{ab} \int_\gamma \mathcal{Y}^{M\hat{a}\hat{b}} \delta_4 \, \ud \tau = \nabla_{ab} \int_\gamma \mathcal{Y}^{M(\hat{a}\hat{b})} \delta_4 \, \ud \tau + \nabla_{[ab]} \int_\gamma \mathcal{Y}^{M\hat{a}\hat{b}} \delta_4 \, \ud \tau$, the first term of which being already in normal form (integrand symmetric with respect to $a$ and $b$ and orthogonal to $u^a$). For the second term we use the definition of the Riemann tensor and its algebraic symmetries to get 
\beq\label{first}
    \nabla_{[ab]} \int_\gamma \mathcal{Y}^{M\hat{a}\hat{b}} \delta_4 \, \ud \tau = -\frac{1}{2} \sum_{j=1}^m \int_\gamma R_{abe}^{\phantom{abe}c_j} \mathcal{Y}^{M_e\hat{a}\hat{b}} \delta_4 \, \ud \tau \, ,
\eeq
where $M_e$ is the multi-index $M$ with $e$ at the $j$-th slot. This term is in normal form since it does not involve any derivative, just like the first term on the right-hand side of Eq.~\eqref{start}. 

Next, for the second term of Eq.~\eqref{troy}, we commute the two covariant derivatives and use once again the definition of the Riemann tensor. Using the formula \eqref{magic} we obtain
\beq \label{second}
\nabla_{ab} \int_\gamma \mathcal{Y}^{Mu\hat{b}}u^a \delta_4 \, \ud \tau = \nabla_{b} \int_\gamma \bigl(\mathcal{Y}^{Mu\hat{b}}\bigr)\,\dot{}\; \delta_4 \, \ud \tau - \sum_{j=1}^m \int_\gamma R_{abe}^{\phantom{abe}c_j} \mathcal{Y}^{M_e u\hat{b}}u^a \delta_4 \, \ud \tau \, ,
\eeq
where the rightmost term is in normal form. However the first term is not, because it needs not be orthogonal to $u_b$. But it can be handled simply by writing the integrand $\bigl(\mathcal{Y}^{Mu\hat{b}}\bigr)\,\dot{}$ as $(\mathcal{Y}^{Muc})\,\dot{}\,h^b_{\phantom{a}c}+\mathcal{Y}^{Muc}\dot{h}^b_{\phantom{a}c}$. The Leibniz rule and metric compatibility imply $\dot{h}^b_{\phantom{a}c}=\dot{u}^b u_c+u^b \dot{u}_c$. We combine these formulas and use the formula \eqref{magic} one last time to get
\beq\label{third_bis}
    \nabla_b \int_\gamma \bigl(\mathcal{Y}^{Mu\hat{b}}\bigr)\,\dot{}\,\delta_4 \, \ud \tau = \nabla_b \int_\gamma \bigl[ \bigl( \mathcal{Y}^{Muc} \bigr)\,\dot{}\,h^b_{\phantom{a}c}
+ \mathcal{Y}^{Muu} \dot{u}^b \bigr] \delta_4 \, \ud \tau 
+ \int_\gamma \bigl(\mathcal{Y}^{Muc}\dot{u}_c\bigr)\,\dot{} \, \delta_4 \, \ud \tau  \, .
\eeq
Finally, for the third and last term of \eqref{troy}, we write, again, an orthonormal decomposition with respect to the abstract index $a$, namely $\mathcal{Y}^{Mau} = \mathcal{Y}^{M\hat{a}u} - \mathcal{Y}^{Muu}u^a$. Taking the covariant derivative along $u^a$ and using the Leibniz rule, along with the formula \eqref{magic}, then gives
\beq \label{third}
\nabla_a \int_\gamma \bigl(\mathcal{Y}^{Mau}\bigr)\,\dot{}\;\delta_4 \, \ud \tau
= \nabla_a \int_\gamma \bigl(\mathcal{Y}^{M\hat{a}u}\bigr)\,\dot{}\;\delta_4 \, \ud \tau
- \nabla_a \int_\gamma \mathcal{Y}^{Muu} \dot{u}^a \delta_4 \, \ud \tau
- \int_\gamma \bigl(\mathcal{Y}^{Muu}\bigr)\,\ddot{}\;\delta_4 \, \ud \tau \, .
\eeq
The second to last term is in normal form since $\dot{u}^a$ is orthogonal to $u_a$, and the last one is in normal form too. Finally, the first term in the right-hand side of Eq.~\eqref{third} can be brought into normal form by following the steps that yielded Eq.~\eqref{third_bis}.

To conclude, we can combine Eqs.~\eqref{first}--\eqref{third} to write the normal form of \eqref{troy}. Combining the latter with \eqref{Deu} gives, at last, the normal form of the quadrupolar expansion \eqref{start} of $Y^M$ according to
\beq\label{start_NF}
    Y^M = \int_\gamma \scY^{M} \delta_4 \, \ud \tau 
+ \nabla_a \int_\gamma \scY^{Ma} \delta_4 \, \ud \tau 
+ \nabla_{ab} \int_\gamma \scY^{Mab} \delta_4 \, \ud \tau \, ,
\eeq
where $\scY^M,\scY^{Ma}$ and $\scY^{Mab}$ are given explicitly in terms of $\mathcal{Y}^M,\mathcal{Y}^{Ma}$ and $\mathcal{Y}^{Mab}$ by
\begin{subequations}\label{scoubidoubidou}
    \begin{align}
    \scY^{M} &= 
      \mathcal{Y}^M - \left( \mathcal{Y}^{Mu} 
    - \bigl(\mathcal{Y}^{Muu}\bigr)\,\dot{}
    + 2\mathcal{Y}^{M(cu)}\dot{u}_c \right)\dot{}\,
    + \sum_{j=1}^m R_{abe}^{\phantom{abe}c_j} \bigl(\mathcal{Y}^{M_e u\hat{b}} u^a - \tfrac{1}{2}\mathcal{Y}^{M_e \hat{a}\hat{b}}\bigr) \, , \\
    \scY^{Ma} &=
      \mathcal{Y}^{M\hat{a}} 
    - 2\bigl(\mathcal{Y}^{M(cu)}\bigr)\,\dot{}\; h^a_{\phantom{a}c}
    - \mathcal{Y}^{Muu}\dot{u}^a \, , \\
    \scY^{Mab} &= \mathcal{Y}^{M(\hat{a}\hat{b})} \, .
   \end{align} 
\end{subequations}
Note that, by construction, $\scY^{Ma}u_a=0$, $\scY^{M[ab]} = 0$ and $\scY^{Mab}u_b=0$. Consequently \eqref{start_NF} is the normal form of \eqref{start}. In particular, this result was used in Sec.~\ref{subsec:Lieconstraints} to go from Eq.~\eqref{Li} to the associated normal form \eqref{Linormal}--\eqref{constraints}. However, the calculations performed above are not sufficient by themselves to derive the reduced form \eqref{SET} of the energy-momentum tensor of a quadrupolar particle, nor the associated equations of evolution \eqref{EE}, which were achieved in Ref.~\cite{StPu.10}. Indeed, while imposing the local conservation law \eqref{DT=0} to the generic quadrupolar energy-momentum tensor \eqref{AnsatzSET}, one must in particular put into normal form the quadrupolar contribution $\nabla_{bcd} \int_\gamma \mathcal{T}^{abcd} \delta_4 \, \ud \tau$, which involves a \textit{triple} covariant derivative.

\section{Collection of results on Killing vector fields} \label{app:Killing}

For the convenience of the reader, we collect here many well-known (and some less known) results for a spacetime $(\mathcal{M},g_{ab})$ endowed with a Killing vector field $\xi^a$. The intrinsic definition of a Killing vector is $\Lixi g_{ab}=0$, which simply states that the metric is invariant along the integral curves of $\xi^a$. Using the metric-compatible covariant derivative $\nabla_a$, this defining equation is equivalent to Killing's equation
\beq\label{Killing}
	\nabla_{(a} \xi_{b)} = 0 \, .
\eeq

\subsection{Compendium of various formulae}\label{subapp:compendium}

We start with a collection of various well-known formulae valid for a generic Killing vector field, that are used throughout this manuscript. First, taking the trace of Killing's equation \eqref{Killing} shows that a Killing vector field $\xi^a$ is divergenceless:
\beq
	\nabla_a \xi^a = 0 \, .
\eeq
Killing's equation also implies the identity $\tfrac{1}{2} \nabla_a (\xi^b \xi_b) = \xi^b \nabla_a \xi_b = - \xi^b \nabla_b \xi_a $ for the ``acceleration'' of a Killing field along its integral curves. Contracting once more with $\xi^a$ shows that the norm squared $\xi^b \xi_b$ of a Killing vector field is conserved along its integral curves:
\beq
	\frac{1}{2} \, \xi^a \nabla_a (\xi^b \xi_b) = \xi^a \xi^b \nabla_{(a} \xi_{b)} = 0 \, .
\eeq

Next, we establish the Kostant formula, which will prove especially useful in the remainder of this Appendix. Combining the defining property of the Riemann tensor with Killing's equation \eqref{Killing} yields
\beq
	\nabla_{ab} \xi_c + \nabla_{bc} \xi_a = R_{abc}^{\phantom{abc}d} \xi_d \, .
	\eeq
Performing a cyclic permutation on the indices $a$, $b$, $c$, and considering the linear combination $(abc) + (cab) - (bca)$, we readily obtain
	\beq
	2\nabla_{ab} \xi_c = \bigl( R_{abc}^{\phantom{abc}d} + R_{cab}^{\phantom{cab}d} - R_{bca}^{\phantom{bca}d} \bigr) \xi_d = - 2 R_{bca}^{\phantom{bca}d} \xi_d \, ,
	\eeq
where the algebraic symmetry property $R_{[abc]}^{\phantom{[abc]}d} \!=\! 0$ was used in the last equality. Finally with $R_{bca}^{\phantom{bca}d} = - R_{cba}^{\phantom{bca}d}$ we obtain the Kostant formula
\beq \label{Kostant}
	\nabla_{ab} \xi_c = R_{cba}^{\phantom{cba}d} \xi_d \, ,
\eeq
or equivalently $\nabla_{ab} \xi^c = R_{dab}^{\phantom{dab}c} \xi^d$. Equation \eqref{Kostant} implies that the 2-form $\nabla_a \xi_b = \nabla_{[a} \xi_{b]}$, as well as its norm $|\nabla \xi|$, are conserved along the integral curves of $\xi^a$. Indeed, by virtue of the antisymmetry of the Riemann tensor with respect to its last two indices,
\beq\label{xiDDxi}
	\xi^c \nabla_c \nabla_a \xi_b = - R_{abcd} \xi^c \xi^d = 0 \, .
\eeq

\subsection{Commutation of the covariant and Lie derivatives}\label{subapp:commutation}
In this subsection we prove that, for any tensor field, the Lie derivative operator $\Lixi$, such that $\Lixi g_{ab} = 0$, commutes with the metric-compatible covariant derivative operator $\nabla_c$, such that $\nabla_c g_{ab} = 0$. First, consider a tensor field $T^N$ of type $(n,0)$, where $N \equiv {c_1 \cdots c_n}$ denotes an abstract multi-index with $n$ indices.\! By definition of the Lie derivative operator\alext{, for the Levi-Civita connection} we have
\begin{subequations}
	\begin{align}
	\Lixi T^N &= \xi^e \nabla_e T^N - \sum_{i = 1}^n T^{N_e} \nabla_e \xi^{c_i} \, , \\
	\Lixi \nabla_a T^N &= \xi^e \nabla_{ea} T^N - \bigl( \nabla^e T^N \bigr) \nabla_e \xi_a - \sum_{i = 1}^n \bigl( \nabla_a T^{N_e} \bigr)  \nabla_e \xi^{c_i} \, , \label{lienab}
	\end{align}
\end{subequations}
where we used the shorthand $N_e \equiv c_1 \cdots e \cdots c_n$, with the abstract index $e$ at the $i$th slot. Taking the covariant derivative of the first equation yields a formula that will be shown to be identical to Eq.~\eqref{lienab}. Indeed,
\beq \label{nablie}
	\nabla_a \Lixi T^N = (\nabla_a \xi^e) \nabla_e T^N + \xi^e \nabla_{ae} T^N - \sum_{i = 1}^n \bigl( \nabla_a T^{N_e} \bigr) \nabla_e \xi^{c_i} - \sum_{i = 1}^n T^{N_e} \nabla_{ae} \xi^{c_i} \, .
\eeq
By Killing's equation, the first term of \eqref{nablie} is identical to the second term of \eqref{lienab}. Since \eqref{nablie} and \eqref{lienab} share the same third term, we get the following expression for their difference:
\beq\label{ALT}
	(\nabla_a \Lixi  - \Lixi \nabla_a) T^N = 2 \xi^e \nabla_{[ae]} T^N - \sum_{i = 1}^n T^{N_e} \nabla_{ae} \xi^{c_i} \, .
	\eeq
Then, we apply the defining property of the curvature tensor to the first term on the right-hand side of \eqref{ALT}, we use $R_{aeb}^{\phantom{aeb}c_i} = - R_{eab}^{\phantom{eab}c_i}$, and rename some indices to obtain
\beq\label{BLT}
	(\nabla_a \Lixi  - \Lixi \nabla_a) T^{c_1 \cdots c_n} = \sum_{i = 1}^n T^{c_1 \cdots e \cdots c_n} (R_{bae}^{\phantom{bae}c_i} \xi^b - \nabla_{ae} \xi^{c_i}) = 0 \, ,
\eeq
where the last equality follows by noticing that each term in parenthesis vanishes, by virtue of Kostant's formula \eqref{Kostant}. Finally, since the metric satisfies $\Lixi g_{ab} = 0$ and $\nabla_c g_{ab} \!=\! 0$, it can be used to ``lower'' indices in \eqref{BLT}, such that the result holds for a tensor field of any type. In summary, we have proven that for any Killing vector field $\xi^a$ and for any tensor field $\bm{T}$,
	\beq \label{Commutation}
	\bm{\nabla} (\Lixi \bm{T}) = \Lixi (\bm{\nabla} \bm{T}) \, .
	\eeq

\subsection{Lie-dragging of various tensor fields}\label{subapp:LieGeo}

Heuristically, we would expect that the Lie-dragging along $\xi^a$ of the metric, i.e. $\Lixi g_{ab}=0$, implies that any tensor field that is constructed geometrically from the metric is Lie-dragged as well. First, we show that the canonical volume form $\varepsilon_{abcd}$ associated with $g_{ab}$ is Lie-dragged along $\xi^a$. Because any 4-form is necessarily proportional to $\varepsilon_{abcd}$, there exists a scalar field $f$ such that $\Lixi \varepsilon_{abcd} = f \varepsilon_{abcd}$. Using $\Lixi g_{ab} = 0$ we thus have
\beq
	\Lixi (\varepsilon^{abcd}\varepsilon_{abcd}) = 2 \varepsilon^{abcd} \Lixi \varepsilon_{abcd} = 2 f \varepsilon^{abcd} \varepsilon_{abcd} \, .
\eeq
The normalization condition $\varepsilon^{abcd}\varepsilon_{abcd} = -4!$ implies that the left-hand side of this equation vanishes, so that $f = 0$, which implies as claimed
\beq\label{Lie-epsilon}
	\Lixi \varepsilon_{abcd} = 0 \, .
\eeq

Second, we prove that the Riemann curvature tensor $R_{abc}^{\phantom{abc}d}$ associated with the metric $g_{ab}$ is Lie-dragged along $\xi^a$. To do so, we consider the Lie derivative $\Lixi \omega_c$ of an arbitrary 1-form field $\omega_c$. Using successively the commutation property \eqref{Commutation}, the defining property of the Riemann tensor and the Leibniz rule for $\Lixi$ gives
	\beq
	2 \nabla_{[a} \nabla_{b]} \Lixi \omega_c = \Lixi (2\nabla_{[a} \nabla_{b]} \omega_c) = \Lixi (R_{abc}^{\phantom{abc}d} \omega_d) = (\Lixi R_{abc}^{\phantom{abc}d}) \, \omega_d + R_{abc}^{\phantom{abc}d} \Lixi \omega_d \, .
	\eeq
By definition of the Riemann tensor the left-hand side of this equation is equal to the second term on the right-hand side, such that  $(\Lixi R_{abc}^{\phantom{abc}d}) \, \omega_d = 0$. Since this equality holds for an arbitrary 1-form $\omega_a$, we conclude that
\beq\label{LieRiem}
	\Lixi R_{abc}^{\phantom{abc}d} = 0 \, .
\eeq
Surprisingly, as far as we know, this important result scarcely appears in classical textbooks, Ref.~\cite{SKMHH} being a notable exception. Now, by using the commutation of the covariant and Lie derivatives, \eqref{LieRiem} implies that any number $n \geqslant 1$ of covariant derivatives of the Riemann tensor is Lie-dragged as well:
\beq\label{LieNablaRiem}
	\Lixi \bigl( \nabla_{e_1 \cdots e_n} R_{abc}^{\phantom{abc}d} \bigr) = \nabla_{e_1 \cdots e_n}  \bigl( \Lixi R_{abc}^{\phantom{abc}d} \bigr) = 0 \, .
\eeq

Finally, we show that if the Einstein equation holds, then the energy-momentum tensor $T^{ab}$ must be Lie-dragged as well. Indeed, combining Eq.~\eqref{LieRiem} with $\Lixi g^{ab} = 0$ readily implies the Lie-dragging of the Ricci tensor $R_{ab} = g^{cd} R_{acbd}$ and of the scalar curvature $R = g^{ab} R_{ab}$. Now recall the Einstein field equation,
	\beq
	R_{ab} - \frac{1}{2} R g_{ab} + \Lambda g_{ab} = 8\pi \, T_{ab} \, ,
	\eeq
where $\Lambda$ is a (cosmological) constant. We established that all the geometrical quantities on the left-hand side of this equation are Lie-dragged along $\xi^a$. Therefore, we conclude that the right-hand side must be Lie-dragged as well, i.e., that
	\beq \label{LieSET}
	\Lixi T^{ab} = 0 \, .
	\eeq
For a given matter source, Eq.~\eqref{LieSET} can be used to constrain the various degrees of freedom encoded in the energy-momentum tensor. As an exemple, let us consider the case of a perfect fluid, for which
	\beq \label{perfectfluid}
	T^{ab} = (\epsilon + p) \, u^a u^b + p \, g^{ab} \, ,
	\eeq
with $\epsilon$ and $p$ the proper energy density and pressure measured by an observer with 4-velocity $u^a$. If the spacetime is endowed with a Killing vector field $\xi^a$, then taking the Lie derivative of Eq.~\eqref{perfectfluid} and the different projections along and orthogonal to $u^a$ yields
\beq
	\Lixi T^{ab} = 0 \quad \Longleftrightarrow \quad
	\begin{cases}
		\;\;\Lixi \epsilon = 0 \, , \\
		\;\,\Lixi p = 0 \, , \\
		\Lixi u^a = 0 \, .
	\end{cases}
\eeq
In the same spirit, for the quadrupolar gravitational skeleton model \eqref{SET} used in this work, we showed in Sec.~\ref{sec:Lips} that $\Lik T^{ab} = 0$ implies the Lie-dragging along the helical Killing field $k^a$ of the components of $T^{ab}$, namely $\Lik u^a = 0$, $\Lik p^a = 0$, $\Lik S^{ab} = 0$ and $\Lik J^{abcd} = 0$, for each particle in the binary system.

\bibliography{ListeRef.bib}

\end{document}